\documentclass[twocolumn]{aastex61}
\usepackage{amsmath,amstext}
\usepackage{graphicx}
\usepackage{CJK}
\usepackage{amsmath}

\newcommand{\beqn}{\begin{equation}}
\newcommand{\eeqn}{\end{equation}}

\begin{document}

\begin{CJK*}{UTF8}{gbsn}

\def\liu#1{{\bf #1}}

\shorttitle{Structure and Color Gradients of UDGs}
\title{Structure and Color Gradients of Ultra-diffuse Galaxies in Distant Massive Galaxy Clusters}
\email{E-mail: fsliu@nao.cas.cn}

\author{Pinsong Zhao}
\affil{School of Astronomy and Space Science, University of Chinese Academy of Sciences, Beijing 100049, China}
\affil{Key Laboratory of Optical Astronomy,
National Astronomical Observatories, Chinese Academy of Sciences, 20A Datun Road, Chaoyang District, Beijing 100101, China}

\author{F. S. Liu $^{\color{blue} \dagger}$}
\affil{School of Astronomy and Space Science, University of Chinese Academy of Sciences, Beijing 100049, China}
\affil{Key Laboratory of Optical Astronomy,
National Astronomical Observatories, Chinese Academy of Sciences, 20A Datun Road, Chaoyang District, Beijing 100101, China}

\author{Qifan Cui}
\affil{Key Laboratory of Space Astronomy and Technology, National Astronomical Observatories, Chinese Academy of Sciences, 20A Datun Road, Chaoyang District, Beijing 100101, China}

\author{Hassen M. Yesuf}
\affil{Kavli Institute for the Physics and Mathematics of the Universe(WPI), UTIAS, University of Tokyo, Kashiwa, Chiba 277-8583, Japan}

\author{Hong Wu}
\affil{School of Astronomy and Space Science, University of Chinese Academy of Sciences, Beijing 100049, China}
\affil{Key Laboratory of Optical Astronomy,
National Astronomical Observatories, Chinese Academy of Sciences, 20A Datun Road, Chaoyang District, Beijing 100101, China}


\begin{abstract}
We have measured structural parameters and radial color profiles of 
108 ultra-diffuse galaxies (UDGs), carefully selected from six distant massive galaxy clusters 
in the Hubble Frontier Fields (HFF) in redshift range from 0.308 to 0.545. 
Our best-fitting GALFIT models show that the HFF UDGs 
have a median S\'ersic index of 1.09, which is close to 0.86 for local UDGs in the Coma cluster. 
The median axis-ratio value is 0.68 for HFF UDGs and 0.74 for Coma UDGs, respectively. 
The structural similarity between HFF and Coma UDGs suggests that they are the same kind of galaxies seen at different times and the structures of UDGs do not change at least for several billion years.
By checking the distribution of HFF UDGs in the rest-frame $UVJ$ and $UVI$ diagrams, 
we find a large fraction of them are star-forming. 
Furthermore, a majority of HFF UDGs show small $\rm U-V$ color gradients within \,1\,*\,$R_{e,SMA}$
region, the fluctuation of the median radial color profile of HFF UDGs is smaller than 0.1\,mag,
which is compatible to Coma UDGs. Our results indicate 
that cluster UDGs may fade or quench in a self-similar way, irrespective of the radial distance, in less than $\sim$ 4 Gyrs.
\keywords{galaxies: photometry --- galaxies: structure --- galaxies: star formation}

\end{abstract}

\section{Introduction}

Decades ago, \citet[][]{SandageBinggeli84} found extremely faint galaxies with unusually large sizes in Virgo. 
After that, more works continued to find low surface brightness dwarf elliptical galaxies in local groups/clusters \citep[e.g.,][]{Thompson+1993,Jerjen+2000,Conselice+2002,Conselice+2003,Mieske+2007}.
Benefiting from the Hubble Space Telescope (HST), people studied the mophologies of these dwarf galaxies in the Perseus cluster and their environmental dependance \citep[e.g.,][]{Penny+2009,deRijcke+2009,Penny+2011}. Dwarf galaxies found in their works show no evidence of tidal process induced by the cluster environment, and their larger petrosian radius indicate that they may have a large dark matter content \citep[][]{Penny+2009}.
These galaxies then attracted a lot of attention in recent years, after \citet[][]{vanDokkum+15a} reported the discovery of 47 Milky Way-sized, extremely diffuse galaxies in their deep imaging survey for Coma cluster using the Dragonfly Telephoto Array, 
they named these galaxies as ultra-diffuse galaxies (UDGs). 
Optical spectroscopic observations have confirmedhave confirmed that some of the UDGs are indeed members of the Coma Cluster \citep[e.g.,][]{vanDokkum+15b,Kadowaki+2017}. 
After that, more and more UDGs are discovered in both cluster 
\citep[e.g.,][]{Koda+15,Mihos+2015,vanderburg+2016,Shi+2017,Venhola+2017,Iodice+2020} and field regions 
\citep[e.g.,][]{Leisman+2017,He+2019,Zaritsky+2021,Kadowaki+2021} in the local Universe from deep imaging survey. 
UDGs in clusters have relatively low sersic indices and red color \citep[][]{Yagi+2016}. 
Some of them host a lot of globular clusters \citep[GCs;][]{van2016,Amorisco2018}; 
spectroscopic observations indicate that most of cluster UDGs have old stellar populations 
and low metallicities \citep[][]{Kadowaki+2017,Gu+2018}. 
In contrast, UDGs in fields or groups seem to have a quite different properties from their counterparts in clusters. 
The field UDGs usually have blue colors and are rich in HI, which indicate that they have ongoing star formation and relatively young stellar population. \citep[][]{He+2019,Trujillo+2017,Rong+2020}.

Because UDGs are extremely diffuse and dim, previous studies could only identify them in the local Universe. However, a sample of UDGs in the distant Universe are needed to study their evolution.
The fartherst UDGs studied are 
by \citet[][]{Bachmann+2021}, who searched for large low surface brightness galaxies 
in two clusters at z = 1.13 and z = 1.23.
Their work showed an under-abundance of UDGs 
in high redshift clusters, by a factor of $\sim$ 3, compared to local clusters. 
The Hubble Frontier Field (HFF) program took deep images of six massive galaxy clusters, 
which provides the best data to study UDGs in distant clusters. 
Several works have already presented the search results of UDGs 
in the HFF and studied their global properties 
\citep[e.g.,][]{Janssens+2017,Janssens+2019,Lee+2017,Lee+2020}. 

Investigating the radial properties of galaxy (i.e., color, star formation rate, etc.) 
is a powerful way to understand how stellar mass is build up and where the star formation is shut 
donw in galaxies \citep[][]{Wu+2005,Liu+2016,Liu+2017,Liu+2018}. 
Works have been done to study the radial stellar population of dwarf elliptical galaxies 
in the local universe \citep[][]{Chilingarian+2009,Koleva+2011},
but there do not have systematic studies on the radial profiles of UDGs.
\citet[][]{Villaume+2022} studied the radial stellar properties of one famous UDG 
in Coma cluster, Dragonfly 44 (DF44), using the Keck Cosmic Web Imager. 
The authors presented evidence that DF44 experienced an intense episode of star formation and then quenched rapidly, unlike canonical dwarf galaxies.
With the aim to understand the assembly and quenching processes in distant UDGs, in this work 
we carefully identify a sample of 108 UDGs in the HFF in redshift range from 0.308 to 
0.545. With this sample, for the first time we make a statistically robust analysis of 
radial color gradients in distant UDGs, and compare their properties 
with the Coma UDGs. 
This paper is organized as follows. 
In Section \ref{sec:data}, we introduce the HFF data and describe how 
we select UDGs and how the imaging processing works. 
In Section \ref{sec:result}, we present the results of our analysis, 
including the global properties of HFF UDGs and their radial color profiles. 
In Section \ref{sec:diss}, we compare our color profiles of HFF UDGs with those of Coma UDGs. 
We also discuss different methods of identifying cluster members and 
describe the effects of distance uncertainties.
Completeness of our UDG sample and a comparison of surface number densities of UDGs among HFF clusters are discussed at last.  
A summary of this work is given in Section \ref{sec:sum}.

Throughout this paper, we adopt a cosmology with a matter density
parameter $\Omega_{\rm m}=0.3$, a cosmological constant
$\Omega_{\rm \Lambda}=0.7$ and a Hubble constant of ${\rm H}_{\rm 0}=70\,{\rm km \, s^{-1} Mpc^{-1}}$.
All magnitudes are in the AB system.

\section{Data} \label{sec:data}

The HFF project is a deep imaging survey, 
which observed 6 massive galaxy clusters--Abell2744, Abell370, AbellS1063, 
MACSJ0416, MACSJ0717 and MACSJ1149--with 6 central cluster field and 6 
coordinated parallel fields by the \textit{HST} ACS/WFC and WFC3/IR cameras for over 840 HST orbits 
\citep[][]{Lotz+2017}. The unprecedented depth of HFF makes it the best data 
to search and study cluster UDGs in the distant Universe. 
For each cluster field, the 30\,mas pixel scale imaging data used in this work 
are collected from the HFF Program in the MAST webpage (\url{https://archive.stsci.edu/prepds/frontier/}), 
which consists of both sci-images and rms-images in the ACS F435W, F606W, F814W bands 
and WFC3 F105W, F125W, F140W, F160W bands. 
These images have been well reduced by the HFF team, 
but extended halos of bright cluster galaxies (bCGs) and diffuse intra cluster light (ICL) 
could be really harmful to studying low surface brightness galaxies.

Fortunately, several works have made efforts in eliminating this effect by modeling 
2-D light distribution of bCGs and ICL and subtract them from original images 
\citep[][]{Castellano+2016,Merlin+2016,Shipley+2018,Pagul+2021}. 
Among them, \citet[][]{Shipley+2018} collected and stacked all existing image data in HFF fields. 
They then reduced the stacked images using a standard procedure, 
including cosmic ray detection, background subtraction, inital source detection, etc. 
After that, bCGs were selected and modeled under an iterative process and finally subtracted from the images. 
On the bCG-subtracted images, they ran {\it SExtractor} \citep[][]{BertinArnouts1996} 
and provided catalogs consisting of total fluxes, flux errors, flux\_radius, 
semi-major/semi-minor axis sizes, etc. 
Sources in the catalog are decteted in a combination of F814W, F105W, F125W, F140W
and F160W bands images, and the F160W band magnitude limits (90\% completeness) range from 26.9\,mag to 27.5\,mag for point sources in deep fields.
In addition, their catalogs also provide photometric 
redshifts (z\_peak) measured using {\it EAZY} code \citep[][]{Brammer+2008}. 
We use all of the above measurements from these catalogs in this work.

\subsection{Selection of UDG candidates} \label{sec:sample}

UDG candidates were selected based on their half-light radii 
and the mean surface brightness within half-light radii. 
By assuming all galaxies in Shipley's photometry catalog \citep[][]{Shipley+2018} are cluster members, 
we first convert their {\it SExtractor} half-light radius, flux\_radius,  into kpc unit 
and compute their mean surface brightness within flux\_radius 
by using following formula

%
%
\begin{equation}
flux\_radius\_kpc =flux\_radius * 0.06 / kpc\_scale
\end{equation}
Here, `flux\_radius' from Shipley's catalog are in pixel unit and the pixel scale is 0.06 arcsec in their work. For different clusters, their redshifts and corresponding kpc\_scale 
values (kpc per arcsec) are listed in Table \ref{tab:table1}. 
\begin{table}
\begin{center}
 \renewcommand\arraystretch{1.0}
 \caption{Redshifts and kpc\_scale of 6 clusters.}
 \label{tab:table1}
 \begin{tabular}{ccc}
  \hline
  Cluster & Redshift & kpc\_scale \\
  \hline
  Abell2744 & $0.308$ & $0.22$ \\
  Abell370 & $0.375$ & $0.194$ \\
  AbellS1063 & $0.348$ & $0.203$ \\
  MACS0416 & $0.396$ & $0.187$ \\
  MACS0717 & $0.545$ & $0.157$ \\
  MACS1149 & $0.543$ & $0.157$ \\
  \hline
 \end{tabular}
\end{center}
\end{table}
%
\begin{equation}
\begin{split}
\langle \mu \rangle _{abs} \ = -2.5*log10(\frac{0.5*flux\_tot}{\pi*(flux\_radius*0.06)^2}) \\
+25-10*log10(1+z\_clu) \\
-Kcorr  ~~~~~~ (mag/arcsec^2)
\end{split} 
\label{eq:mu}
\end{equation}
here, `flux\_tot' from Shipley's catalog is the total flux of galaxy,
`25' is the zeropoint used in their catalog. 
For galaxies in different HFF fields, we correct their cosmic dimming effects 
by using redshifts of clusters listed in Table \ref{tab:table1}.
It is noted that the {\it SExtractor} flux\_radius is a 
rather poor proxy for the true half-light radius and without proper estimate for the 
S\'ersic index \citep[][]{Barden+2012}. The initial use of flux\_radius is to conservatively select all objects large enough
to be a UDG candidate since the observed, PSF-smeared flux\_radius values are larger than 
the true half-light radii for our galaxies. We describe the determination of 
the intrinsic half-light radii of selected candidates in Section 2.3.

We here use the parameters measured in F814W band since 
the observed F814W band is closer to 
the rest-frame SDSS r-band for galaxies at z = 0.3-0.5. We adopt similar UDG selection criteria, 
namely $\langle \mu_{F814W} \rangle _{abs}$ \,$\textgreater\,24 mag/arcsec^2$ and 
flux\_radius\_kpc\,$\textgreater$\,1.5\,kpc, as \citet[][]{Yagi+2016}.
In order to use this criteria, we also include a K-correction term in Eq.\ref{eq:mu}.
For galaxies with redshift at $z\sim0.3$, by simply treating SDSS r-band as a blue shift of F814W band, this term could be written as $Kcorr=-2.5*log_{10}(1+z\_clu)$, 
which is independent of the shape of SED of galaxies \citep[][]{hogg+2002,blanton+2003}. 
For galaxies in Abell\,2744 field, $Kcorr$ equals to $-0.29\,mag$, 
For galaxies in other HFF clusters, we also take $Kcorr$ as $-0.29\,mag$ considering there will be a magnitude difference 
between observed F814W band and $r^{z\_clu}$ band ($r^{z\_clu}$ band is referring to a red-shifted SDSS r-band to redshift $z=z\_clu$).
We also apply a photometric redshift cut. 
The typical uncertainty of photometric redshifts is $\sigma_z$\,$\sim$\,0.03.
Though this uncertanty will increase to $\sim$\,0.3 for objects with F814W magnitudes fainter than 25\,mag,  
in this work, we use a narrow redshift cut, $|$z\_peak$\,-\,$z\_clu$|\,<0.1$,
which helps us effectively remove the background and foreground contaminants. (see Section 4.2 for 
discussions).
We then visually inspect every candidate that satisfies the above criteria. 
Galaxies which have bright neighbors/companions or are located near the edge of the images are rejected. 
Finally, we select out 285 candidates in 6 HFF cluster fileds.


\subsection{Imaging Processing} \label{sec:imgprocess} 

For each UDG candidate, we cutout images in all bands with sizes of $1000\times 1000$ pixels and centers are at the location of the UDG. 
These images are then convolved to have the same 
point spread function (PSF) as those observed in F160W band. 
Due to the extremely low surface brightness of UDGs, it is important to apply a careful
background subtraction before we do accurate analysis of radial light profiles 
\citep[][]{Liu+2016}.
We first run a source detection script using `Noisechisel' \citep[][]{Akhlaghi+2015}, 
which has been tested to have a powerful ability in detecting low-level signals from the noise 
\citep[][]{Haigh+2021}. After masking all pixels hosting signals, we use a median filtering 
to build background images from unmasked background pixels.
The size of median filtering window is flexible from 31 to 251 pixels, , 
depending on the size of each candidate galaxy.
The background images are then 
subtracted from PSF-matched cutout-images. The median value of background reduces 
by 90\% after our background subtraction.

\begin{figure*}[!ht]
\centering
\includegraphics[scale=0.6]{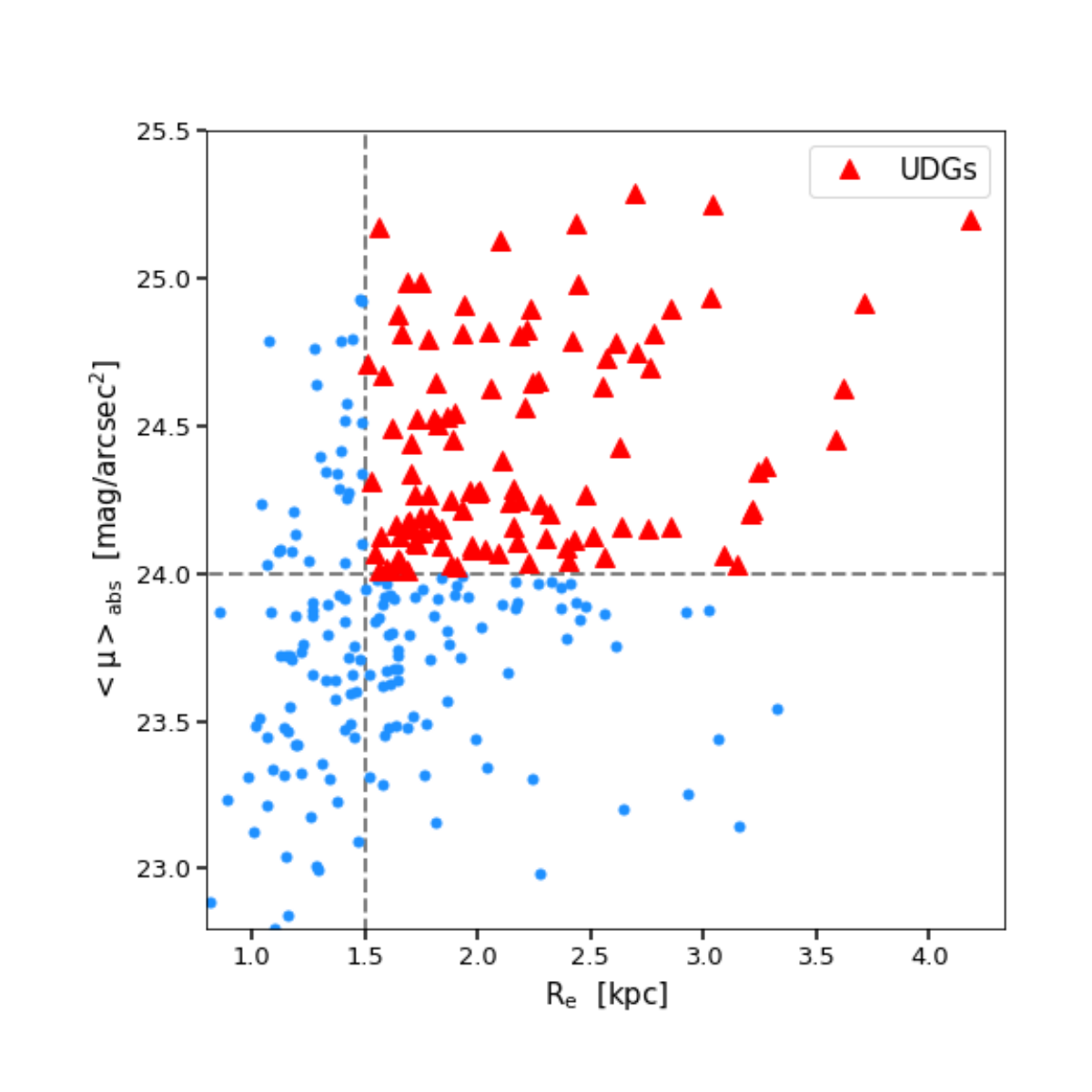}
\caption{Mean surface brightness within circularized effective radius versus effective radius for all UDG candidates.
In total, 108 UDGs are selected from the upper-right region, which are marked as red triangles.
Blue circles are the rest of candidates.}
\label{fig:sizesb}
\end{figure*}

\subsection{GALFIT Fitting and Final Selection of UDGs}

We use GALFIT \citep[][]{Peng+2002,Peng+2010} to fit the single S\'ersic model 
\citep[][]{Sersic1968} to each candidate. The fitting is done on $151\times151$ pixels F814W images, 
which have been background-subtracted as described in Sec \ref{sec:imgprocess}, 
but not PSF-matched. 
Before running galfit, we take use of the detection image from Noisechisel and the segmentation image from SExtractor to mask contamination pixels in the fields.
This could help us to get robust fitting results.
To avoid unreasonable fits, we restrict the ranges of S\'ersic  indices to be between 0.2 and 8 and those of the effective radii to be within 0.5 to 50 pixels. 
Examples of our fits are presented in Appendix.

Based on the best-fitting parameters from GALFIT models, 
we re-determine the effective radii and mean surface brightness of 
285 UDG candidates more accurately.
The distribution of their surface brightness versus radius are shown in Fig.\,\ref{fig:sizesb}. 
Finally, 108 candidates are confirmed to be our UDGs and are marked as red triangles in the top-right region of Fig.\,\ref{fig:sizesb}.

For reference,
Table \ref{tab:table2} lists the numbers of galaxies in each field after we apply different cuts to Shipley's catalog.
 as well as the fraction of galaxies compared to the number of all detected sources in each cluster field.

\begin{table*}
\begin{center}
 \renewcommand\arraystretch{1.0}
 \caption{Numbers of galaxies after different cuts in this work.}
 \label{tab:table2}
 \begin{tabular}{cccccc}
  \hline
  Cluster & total & SB\&Re cut (SEx-based) & photo-z cut & Visually Check & SB\&Re cut (Galfit-based) \\
  \hline
  Abell2744  & $9390$ & $1872(19.94\%)$ & $263(2.80\%)$ & $56(0.60\%)$ & $26(0.28\%)$ \\
  Abell370   & $6795$ & $1577(23.21\%)$ & $163(2.40\%)$ & $63(0.93\%)$ & $23(0.34\%)$\\
  AbellS1063 & $7611$ & $1726(22.68\%)$ & $221(2.90\%)$ & $80(1.05\%)$ & $36(0.47\%)$\\
  MACS0416   & $7431$ & $1600(21.53\%)$ & $145(1.95\%)$ & $37(0.50\%)$ & $9(0.12\%)$\\
  MACS0717   & $6370$ & $1460(22.92\%)$ & $136(2.14\%)$ & $21(0.33\%)$ & $6(0.09\%)$\\
  MACS1149   & $6868$ & $1715(24.97\%)$ & $110(1.60\%)$ & $28(0.41\%)$ & $8(0.12\%)$\\
  \hline
 \end{tabular}
\end{center}
\end{table*}

\section{Results}  \label{sec:result}

\subsection{Structural Properties}

\begin{figure*}[!ht]
\centering
\includegraphics[scale=0.4]{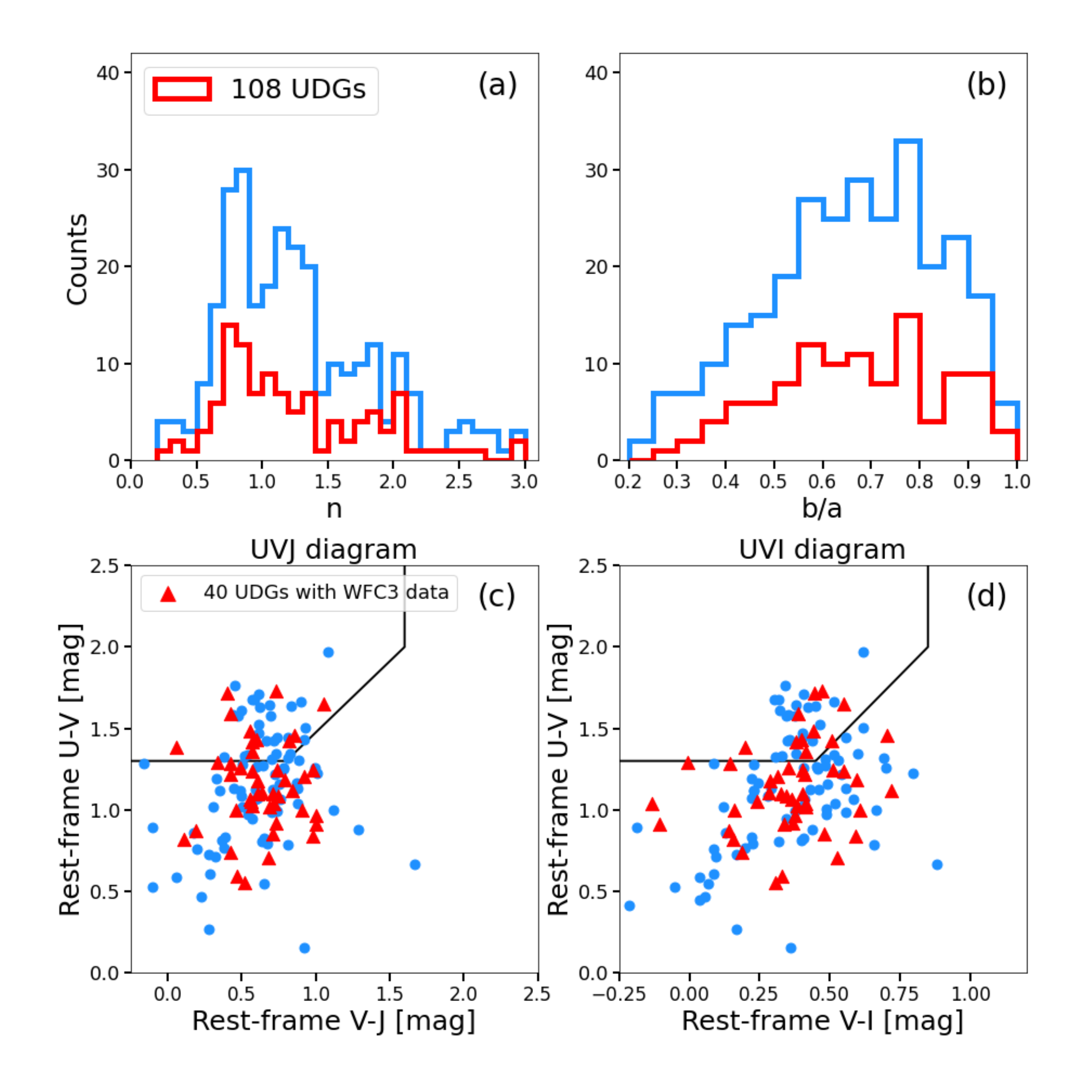}
\caption{Statistics of basic properties of UDGs in this work.
Panel\,(a) and (b) show histograms of the best-fitting S\'ersic index n and b/a, respectively.
UDGs are in red and all candidates are in blue.
In Panel\,(c) and (d), we show the rest-frame UVJ and UVI diagrams for
objects with HST WFC3/IR observations,
whose rest-frame colors are obtained from the EAZY.
In this work, 129 of 285 candidate galaxies and 40 of 108 UDGs have WFC3/IR data.
}
\label{fig:prope}
\end{figure*}

The histograms of S\'ersic index and axis-ratios of the best-fitting GALFIT models are presented 
in panel\,(a) and panel\,(b) in Fig.\,\ref{fig:prope}, blue for all candidates while red for UDGs. 
Similar to UDGs in the local universe, UDGs in HFF fields have relative small n values and 
are not preferentially edge-on galaxies. In this work, 69\% of UDGs have n smaller than 1.5 and 
the median value of n is 1.09, 82\% of UDGs have b/a larger than 0.5 and the median value of 
b/a is 0.68. The statistics of Coma UDGs are 0.86 for n and 0.74 for b/a \citep[][]{Yagi+2016}. 
This means that current structural characteristics of local UDGs might be shaped earlier 
than z$\sim$0.4 and remain unchanged after a relatively long time ($\sim$ 3--5\, Gyrs). 

For galaxies having HST WFC3/IR coverage in HFF project, 
we run EAZY to get their rest-frame UVIJ magnitudes. 
Multi-band fluxes input to EAZY are based on our PSF-matched and background-subtracted images, 
input redshifts are fixed to be cluster redshifts. 
Rest-frame UVJ and UVI diagrams are shown in the panel\,(c) and 
panel\,(d) of Fig.\,\ref{fig:prope}, respectively.
In our sample, 91 candidate galaxies and 14 UDGs have HST WFC3/IR data,
they are marked as blue circles for candidate galaxies and red triangles for UDGs. 
Unlike cluster UDGs in the local Universe, which are found to be red in color and 
show no evidence of recent star forming activities, 
most of cluster UDGs in HFF populate the lower left region of UVJ and UVI diagrams in Fig.\,\ref{fig:prope},
which indicate that they are still star-forming.
%
The relative blue rest-frame V\,-\,J colors ($V-J<1$) indicate that these UDGs contain 
small amount of dust content. 
By assuming all of our UDGs have relative small V\,-\,J color, we 
use rest-frame U\,-\,V color to investigate their star formation activity. 
Utilizing calibrations between rest-frame U\,-\,V and observed F606W\,-\,F814W in \citet[][]{Wang+2017},
we calculate the rest-frame U\,-\,V colors for all 108 UDGs from their observed colors.

\subsection{Radial Color Profiles}

\begin{figure*}[ht]
\centering
\includegraphics[scale=0.2]{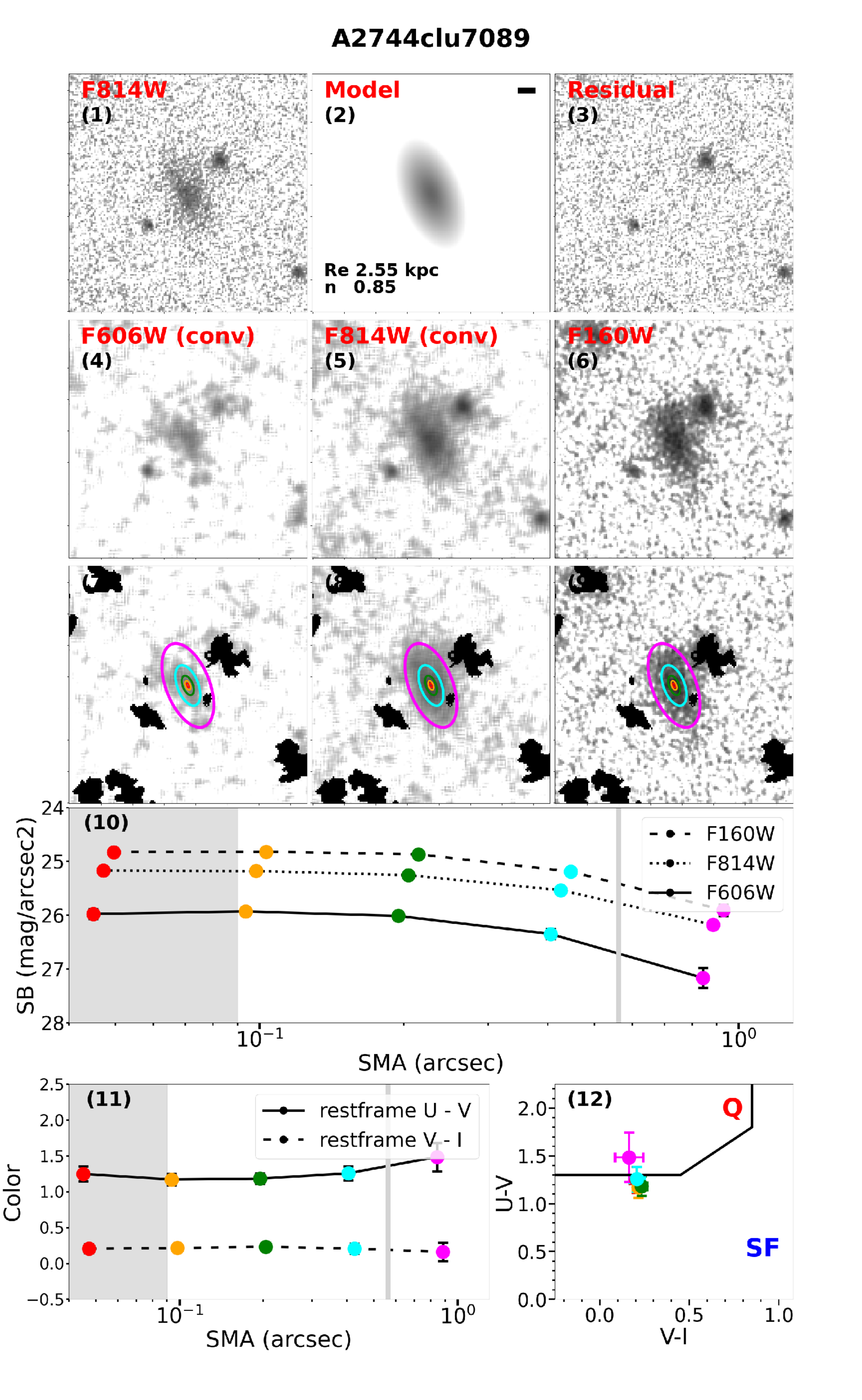}
\caption{Example multi-band surface bright-
ness profiles fitting for UDG A2744clu7089.
Panels\,(1) to (3) show the F814W band cutout-images of the UDG, the best-fitting GALFIT model and residual image. The bar at top-right of panel\,(2) represents 1.5\,kpc assuming the cluster redshift.
Panels\,(4) to (6) show PSF-matched images in F606W, F814W and F160W bands.
In panels\,(7) to (9), we mask neighboring sources classified by `Noisechisel' and overplot our elliptical annuli used in surface brightness analysis.
Panel\,(10) presents three-band surface brightness profiles of UDG A2744clu7089,
x-axis of colorful points correspond to the out-radius of elliptical annuli.
In panel\,(11), we convert observed color profiles into rest-frame U\,-\,V and V\,-\,I profiles.
F814W and F160W surface brightness profiles in panel\,(10) and rest-frame V\,-\,I profiles in panel\,(11) are shifted a bit to the right.
Finally, the colors of the UDG from inside to outside are shown in the UVI diagram in panel\,(12).}
\label{fig:a27447089}
\end{figure*}

For each UDG, we estimate the mean surface brightness within a sequence of elliptical annuli. 
The parameters of annuli are taken from the best-fit GALFIT model of a UDG and 
they are applied to every band, During the computation, we fix the parameters for annuli 
from inside to outside. In Fig.\,\ref{fig:a27447089}, there is an example show our results 
on multi-band surface brightness profiles of UDGs. Panels\,(7) to (9) are PSF-matched 
background-subtracted cutout-images in F606W, F814W and F160W bands. The surface brightness 
in each band is calculated within the region between two neighboring colored ellipses. 
The final surface brightness profile is presented in panel\,(10). In panel\,(11), referring to Fig.\,A1 and Fig.\,A2 in 
\citet[][]{Wang+2017}, we obtain the rest-frame U\,-\,V and V\,-\,I color profiles 
from observed F606W\,-\,F814W and F814W\,-\,F160W profiles, empirically. The shaded regions 
in panel\,(10) and panel\,(11) indicate the half of FWHM of PSF in F160W band, 
and the gray lines show the effective radii along semi-major axis. 
In panel\,(12), we exhibit rest-frame U\,-\,V versus V\,-\,I colors for all annuli. 
Figures for other 13 UDGs with WFC3/IR data are presented in the Appendix.
Similar figures for a full version of all 108 UDGs could be found here (\url{https://drive.google.com/file/d/1dmYcVn0zOi07R4WOXbxh7yNZ_GTKK623/view?usp=sharing}).

It can be seen that these HFF UDGs do not show significantly large color gradients 
within their effective radii, except for UDG AS1063clu2960. Meanwhile, there is a 
large fraction of UDGs that are undergoing star formation activities from inside to outside. 
These findings suggest that UDGs in distant clusters generally grow at a uniform rate throughout the galaxy.

%
%

\begin{figure*}[!ht]
\centering
\includegraphics[scale=0.38]{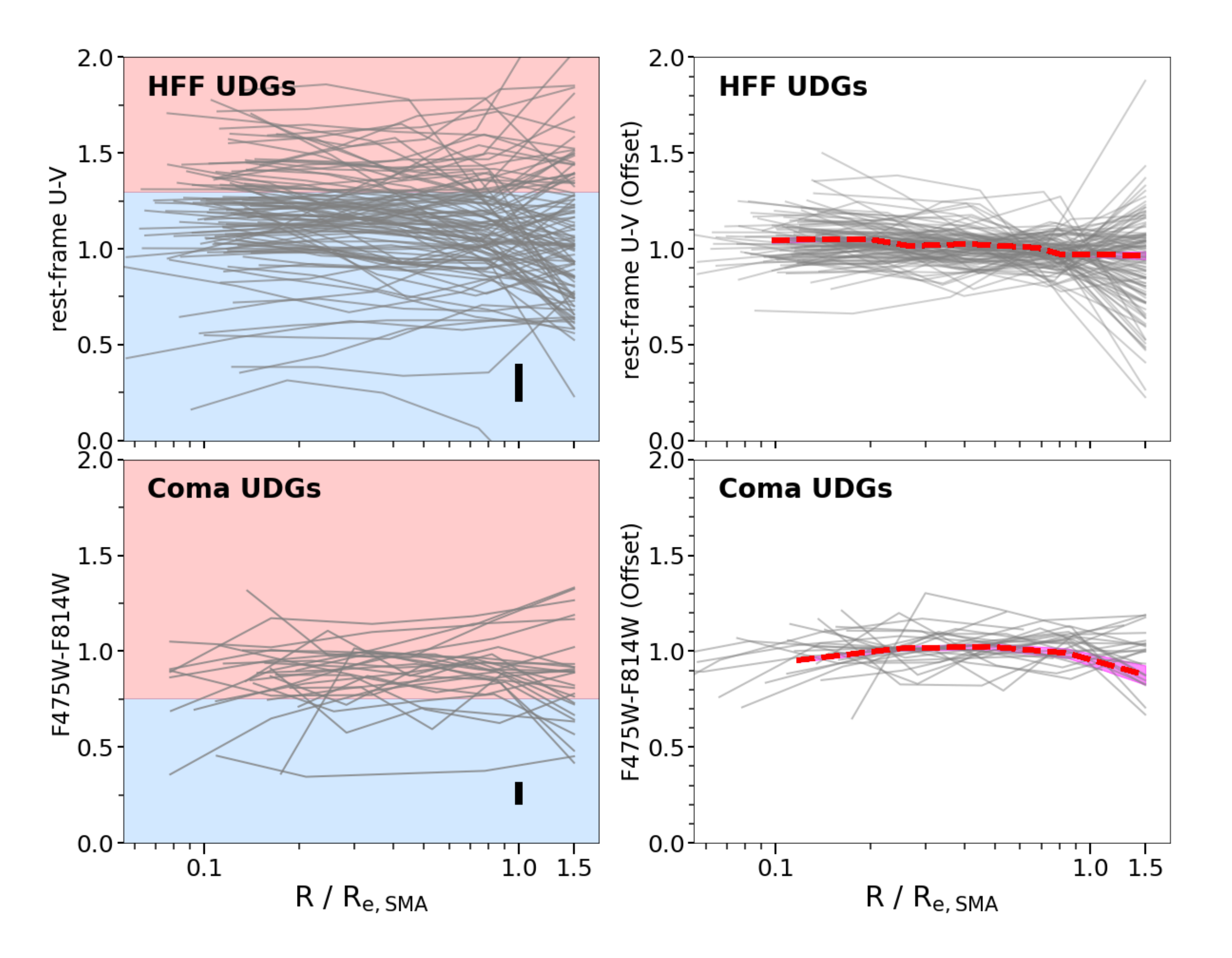}
\caption{The radial color profiles of HFF UDGs and Coma UDGs.
In the left two panels, the rest-frame U\,-\,V color profile of 108 HFF UDGs and HST F475W\,-\,F814W color profiles of Coma UDGs are presented, respectively.
Blue and red backgrounds are used to show the classical separation for blue and red galaxies,
the boundary we used here is 1.3\,mag for the rest-frame U\,-\,V color and 0.76\,mag for F475W\,-\,F814W color.
Here 0.76\,mag is the traditional g\,-\,i color of cluster UDGs, Coma UDGs are known to be red and quenched, 
our F475W\,-\,F814W profiles show that, Coma UDGs are red from inside to outside and have little color gradients. 
The typical uncertainty of the color near $\sim$\,1\,*\,$R_{e,SMA}$ is 0.2\,mag for HFF UDGs and 0.12\,mag for Coma UDGs, which are shown as black bars, respectively.
In the right two panels, we offset each color profile in the left two panels by letting its data points have an average value equal to 1.
The median curve and corresponding 1-sigma uncertainty for the shifted color profiles is shown as red dashed line.
}
\label{fig:profile}
\end{figure*}

\section{Discussions}  \label{sec:diss}

\subsection{Comparison with UDGs in the Coma Cluster} 

Lots of UDGs have been identified in Coma cluster, these Coma UDGs are found to be red and 
have old stellar population. 
Benefiting from the HST/ACS Coma Cluster Treasury Survey \citep{Carter+2008,Hammer+2010}, 
which provides deep and high resolution images in F475W 
and F814W bands, we do similar analysis for Coma UDGs selected from \citet[][]{Yagi+2016} catalog.
Since the Coma UDGs usually have much larger angular size than the size of HST PSF, 
we do not match cutout-images to have the same PSF but only carefully subtract the background, 
set the $SMA$ of the innermost annulus for Coma UDGs are always begin at r\,=\,7\,pixel, 
beyond which we need not worry about the PSF effect.

The comparison of color profiles between HFF UDGs and Coma UDGs are shown in Fig.\,\ref{fig:profile}. 
For HFF UDGs, in left panel, we present the rest-frame U\,-\,V profiles of all 108 UDGs. 
In right panel, we offset each U\,-\,V profile by a mean distance of all annuli colors 
from `y\,=\,1', the median curve are plotted as red dashed line, magenta region 
show the 1--sigma uncertainties. For Coma UDGs, we do similar analysis on F475W\,-\,F814W profiles. 
Within the range of $0.1*R_{e,SMA}$ to $1.5*R_{e,SMA}$, 
both HFF UDGs and Coma UDGs have very small color gradients, 
the changes in color are smaller than 0.1 magnitude. 
Combining the lack of color gradients in both samples with the fact that two samples have very different colors (are starforming and quenched) indicates that cluster UDGs may fade or quench in a self-similar way 
in less than $\sim$ 4 Gyrs.


\begin{figure*}[!ht]
\centering
\includegraphics[scale=0.25]{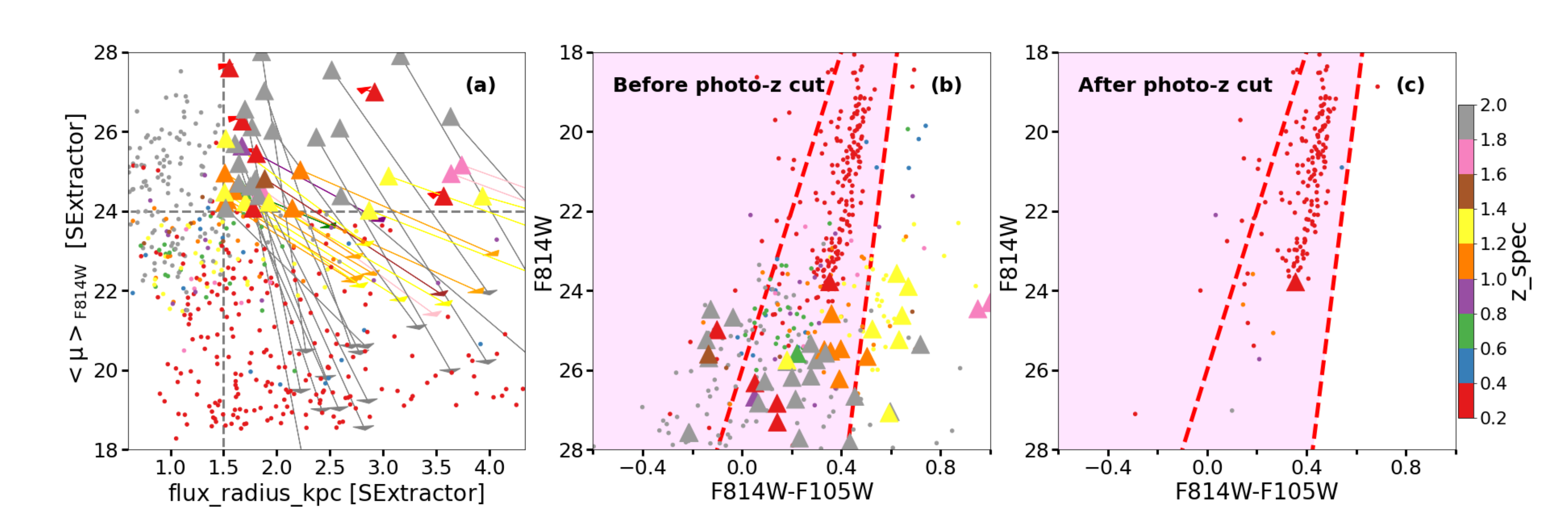}
\caption{
Panel\,(a) shows objects which have spectroscopic redshifts in Abell\,2744 cluster field.
They are plotted in the mean surface brightness versus effective radius space.
Their surface brightness and radius are calculated by assuming the cluster redshift, using parameters from SExtractor.
Triangles in the upper right region represent galaxies having low surface brightness and extened sizes under such assumption, their true values of surface brightness and half-light radius are shown with the arrow.
In panel\,(b), the same sample is plotted in the F814W versus F814W\,-\,F105W space, two red dashed lines show boundaries of the 'red sequence', the shaded region is where \citet{Lee+2017,Lee+2020} select their final UDGs.
Panel\,(c) is a copy of panel\,(b), but applying the photo-z cut we used in this paper for Abell\,2744 cluster.
Each object is colored by their spectroscopic redshift.
}
\label{fig:discussion1}
\end{figure*}

\subsection{Accuracy of Cluster Member identification} 

One of the biggest problems in identifying distant UDGs is their redshift/distance information, 
without which it is difficult to correctly determine their 
absolute magnitudes, unbiased surface brightness correction for the cosmic dimming effect, 
physical sizes, etc..
%

Since it has been known that UDGs in fields have quite different star formation activities 
from UDGs in clusters in the local universe, it would be important to select a relatively 
clean sample of UDGs in distant clusters with less background and foreground objects.
Although spectroscopic observations are the most secure way to determine distances 
and cluster membership, they are too expensive to work for a large sample of distant UDGs. 
For instance, \citet{Kadowaki+2021} reported that $\sim$\,1 hour exposure time on 10\,m class 
telescopes often fails to yield a redshift for a candidate UDG in the Coma region.
Previous studies utilized the color-magnitude diagram, \citet{Lee+2017,Lee+2020} kick out candidates which have color redder than the 'red sequence' of bright cluster galaxies to 
get rid of background sources.
This method definitely helps to remove background candidates from the sample,
but how good is it?

In panel\,(a) of Fig.\,\ref{fig:discussion1}, we show the distribution 
of galaxies, which have spectroscopic redshifts (z\_spec) in Abell\,2744 cluster field, 
in the surface brightness versus radius space. 
Spectroscopic redshifts of these galaxies are taken from \citet{Shipley+2018}, 
which come from five liternature catalogs (see their Section 5.1 for details).
The surface brightenss and half-light radii of these galaxies are calculated 
using the SExtractor parameters, by assuming they have the same redshift as the target cluster, 
just like what previous works did to select UDG candidates 
\citep{Janssens+2017,Janssens+2019,Lee+2017,Lee+2020}. 
Galaxies in the upper-right region of panel\,(a) have surface brightness fainter than 
24\,$mag/arcsec^2$ in this way, but the true surface brighntness determined with their 
spec-z are much brighter and are shown with arrows. 
In panel\,(b) of Fig.\,\ref{fig:discussion1}, 
we show all z\_spec-confirmed objects in the F814W versus F814W-F105W space, 
red dashed lines show the boundaries of the `red sequence', which are used in \citet{Lee+2017}. 
It can be seen that simply removing objects redder than the `red sequence' 
can help to remove some background galaxies, but the remaining sample 
still suffer from the contamination of a large fraction of interlopers.
In panel\,(c) of Fig.\,\ref{fig:discussion1}, we present the result after applying our photo-z cut 
to this sample. It is clearly that, with the help of photometric redshifts, 
the majority of backgroud interlopers can be removed successfully,
It should be noted that in this work we utilize photometric redshifts 
to effectively remove the background and foreground contaminants, but meanwhile, the sample size reduces.
However, the sample purity is critically important for this study. 
In addition, applying a narrow range of photo-z cut is also crucial to correctly 
estimate the mean surface brightness and physical radius of sample galaxies, otherwise, 
the results could be far from the truth, as shown in panel\,(a) of Fig.\,\ref{fig:discussion1}.

We also utilize the XDF as a comparison to estimate the potential contributions of 
field UDGs or background galaxies to our cluster UDG sample. 
The XDF area we used cover $\sim 11$ arcmin$^2$, which are 
downloaded from the webpage \url{https://archive.stsci.edu/prepds/xdf/}.
Photometric redshifts we used are from CANDELS team \citep{Santini+15}.
We re-do our UDG selection process for the XDF. 
Referring to the redshift of each HFF cluster, we apply the same redshift cut for XDF galaxies. 
As a result, no UDGs are found in the XDF for the redshift of M0416, one UDG is found for the redshifts of Abell2744, Abell370 and AS1063,
while two UDGs are seen in the XDF for the redshifts of M0717 and M1149.
The findings indicate that the number density of potential interlopers in 
our selected UDGs from the six HFF clusters is very low ($0/0.1/0.2$ per arcmin$^2$ 
for one specific field).


%

\subsection{Uncertainties in Photometric Redshifts}

\begin{figure*}[!ht]
\centering
\includegraphics[scale=0.5]{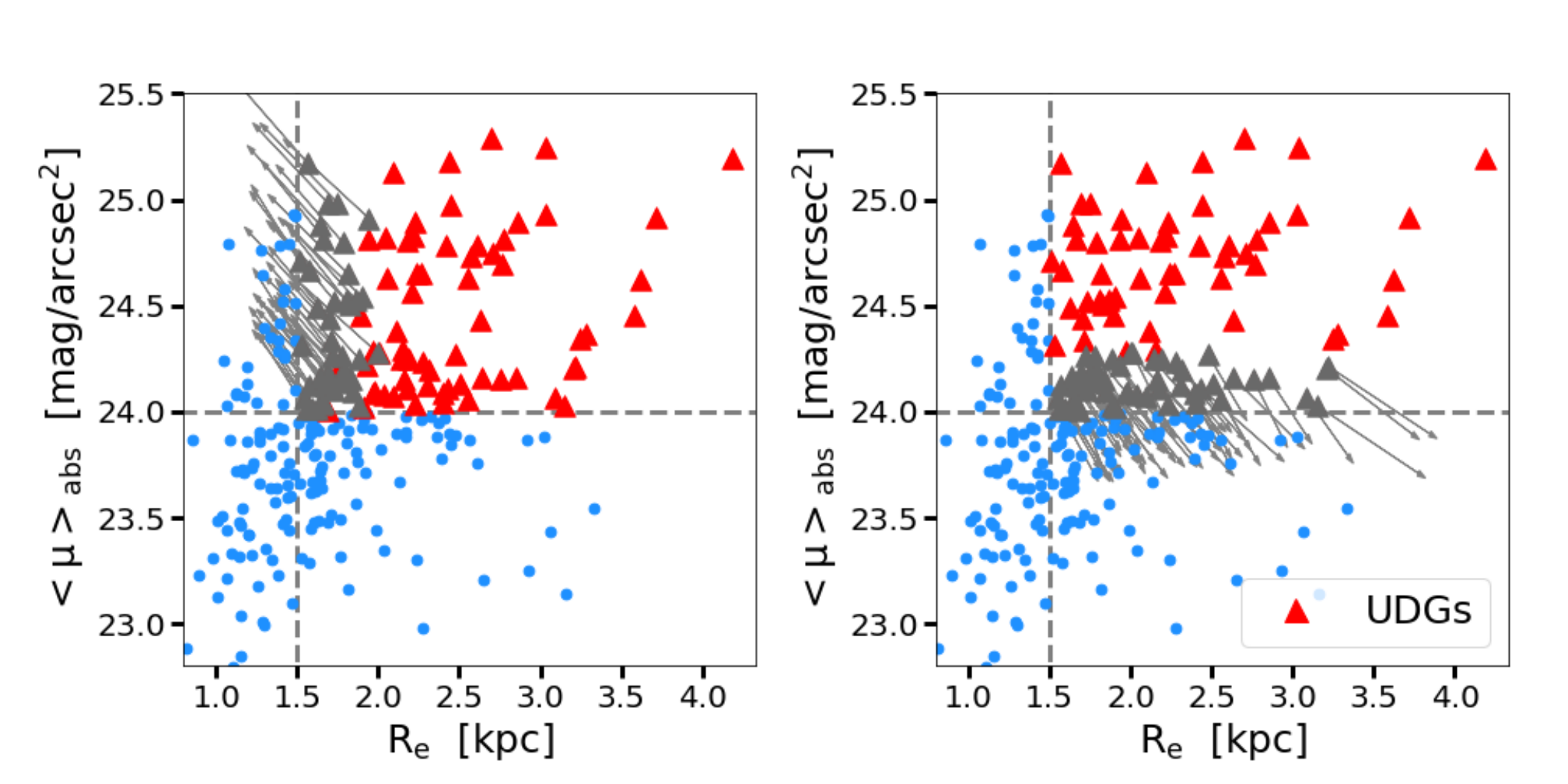}
\caption{The diagram demonstrate the uncertainties in our sample selection.
In left panel, we marked UDGs with gray triangles which will not be classified as UDGs when they are at redshifts equal to z\_clu-0.1.
The right panel show cases when UDGs are assumed to have their redshifts equal to z\_clu-0.1.}
\label{fig:discussion2}
\end{figure*}

In this work, we use photometric redshifts to select cluster members and our UDG sample,
but the uncertainties in photometric redshifts produce uncertainties in surface brightness and physical radius of galaxies,
which will make UDGs move out from the upper-right space in Fig.\ref{fig:sizesb}.
In this section, we simply evaluate this effect.
In Sec.\ref{sec:sample}, we apply $|$z\_peak$\,-\,$z\_clu$|\,<0.1$ to select cluster members.
Referring to \citet{Shipley+2018}, 
around 80\% of our candidate UDGs are located within this redshift range.
But $\pm$\,0.1 uncertainty in redshift is not so accurate,
which would introduce large uncertainties in the estimations of both surface brightness and radius when we select candidate UDGs.
We re-estimate the surface brightness and radius of our UDGs under two conditions,
assuming they have redshifts equal to z\_clu-0.1 or z\_clu+0.1.
Objects which will not be classified as UDGs are marked with gray triangles in two differnt panels in Fig.\ref{fig:discussion2}, separately.
The gray arrows show where they will go.
Two thirds of galaxies in our UDG sample will still satisfy the definition for UDGs if we assume a redshift of z\_clu-0.1, and half of our UDGs will survive for z\_clu+0.1.
The uncertainty in photometric redshifts and the resulting changes in UDG sample sizes 
do not influence our main conclusions.

\subsection{Completeness of our UDG sample}
We run image simulations to evaluate the completeness of our UDG sample.
Firstly, we use GALFIT to generate mock images in F814W band for each cluster, each mock image has a size of 151x151 pixels. 
Model parameters are chosen in the following way:
S\'ersic index is fixed to be n\,=\,1. 
Circularized half-light radius is randomly chosen from a uniform distribution with a range of $1.5<R_{e}<7.5$ kpc.
Total magnitude is randomly chosen from a uniform distribution with a range of $22<mag<29.5$ mag.
Axis-ratio is set to follow a gaussian distribution with mean value of 0.7 and scatter value of 0.1, 
axis-ratio values lager than 1 or smaller than 0.1 will be fixed to be 1 or 0.1, respectively.
As for position angle values, we just choose from 0-360 degrees, randomly.
We generate 10000 model galaxies for each cluster and compute their absolute surface brightness as the same way we did for observed data.
Only mock images who have $24< \langle \mu \rangle_{abs} <28$ mag/arcsec$^{2}$ will be used for the next step.
In general. we will have $\sim$6400 mock UDGs for each cluster. 

Secondly, we inject these mock UDGs into HFF F814W band images.
To avoid an overcrowding in the simulations, we randomly pick up $\sim$40 mock galaxies from mock galaxy sample each time.
For each cluster field, we run 500 simulations.    
We take use of the segmentation image to reduce the posibility of overlaping with other sources.

\begin{figure*}[!ht]
\centering
\includegraphics[scale=0.4]{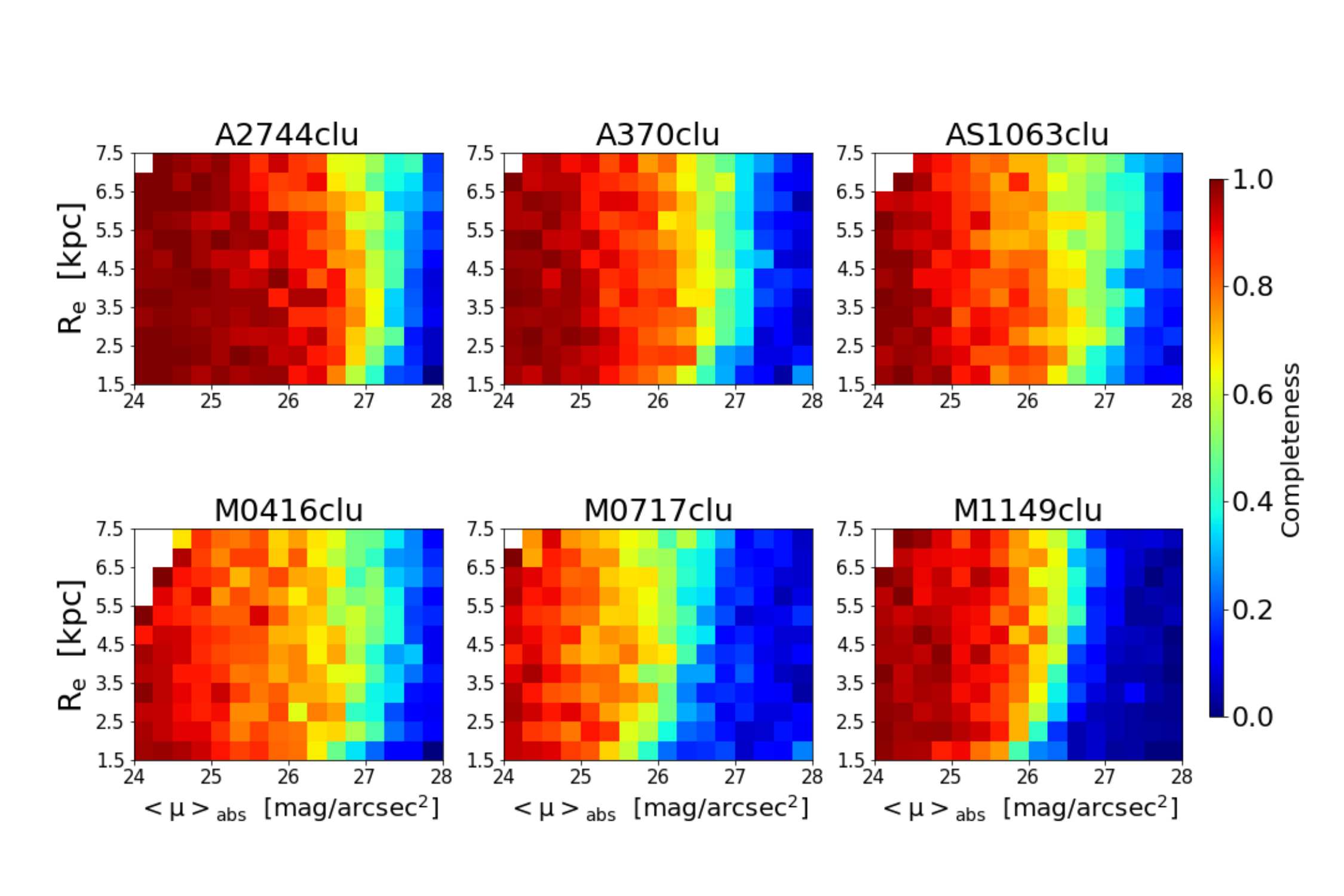}
\caption{Completeness maps as a function of size and surface brightness for n=1 S\'ersic profiles.}
\label{fig:completeness}
\end{figure*}

Lastly, we use SExtractor to detect these mock UDGs, an matching radius of 3 pixel is applied.
The completeness map of absolute surface brightness versus. effective raidus for each HFF cluster is shown in Fig.\,\ref{fig:completeness}.
The completeness here is defined as a ratio of the number of detected UDGs to the number of injected UDGs.
For UDGs with surface brightness brighter than 25.3 mag/arcsec$^{2}$ (the dimmest UDG in this work has a surface brightness of 25.3 mag/arcsec$^{2}$), the completeness in all six clusters are better than 80\%.

\subsection{Concerns of UDG sample size compared with previous works}

\citet[][]{Lee+2017,Lee+2020} identified 27 UDGs in A2744 cluster field, 
34 UDGs in A370 cluster field and 35 UDGs in AS1063 cluster field, 
whereas the numbers of UDGs we found from these fields 
are 26,23 and 36, respectively. 
Two works found similar numbers of UDGs from these three clusters.
\citet[][]{Janssens+2019} found more UDGs than our work, and in particular 
the numbers of UDGs they found in three more distant clusters, MACS0416, MACS0717 and MACS1149, are comparable 
to those in other three clusters.
However, the number of UDGs we found from MACS0416, MACS0717 and MACS1149 are far less than the other three clusters.
To check this, we especially loosen the selection criteria of UDG candidates in Section 2.1 to
be $\langle \mu_{F814W} \rangle$ \,$\textgreater\,22.5 mag/arcsec^2$ and
flux\_radius\_kpc\,$\textgreater$\,1.0\,kpc, and re-do our sample selection process.
As a result, the sample size of final UDGs in the six cluster fields increases to 131,
of which four are in MACS0416, four in MACS0717 and two in MACS1149.
Main results and conclusions we have drawn remain unchanged when 
using this enlarged sample of candidate UDGs.

The differences of sample sizes between this study and previous works 
primarily result from applying a narrow photo-z cut during our selection 
process. 
In \citet[][]{Lee+2017, Lee+2020}, they did not apply any photo-z cut 
to their sample. 
In \citet[][]{Janssens+2019}, they just restricted their candidates to have photo-z less than 1. 
We cross-match selected 27 UDGs in A2744 by Lee et al. with the catalog 
of \citet{Shipley+2018} to obtain 25 UDGs with photo-z measurement.
If we apply the same photo-z cut used in this work to these 25 UDGs of Lee et al, 
only 8 UDGs will pass the photo-z cut. 
The result indicates that the UDG sample of Lee et al may be more affected by foreground/background interlopers than initially thought.

%

It has been proposed that UDGs are born primarily in the field, later processed in groups and, 
ultimately, infall into galaxy clusters \citep[e.g.,][]{Roman+2017}.
Using large simulations, \citet[][]{Rong+2017} found that UDGs could be a type of dwarf galaxies 
residing in the low-density regions hosted by large spin halos, which 
fell into the clusters with a median infall time of $\sim$8.9\,Gyr, 
corresponding to a redshift of 0.43. 
\citet[][]{Tremmel20} also showed that UDGs in cluster environments form 
from dwarf galaxies that experienced early cluster in-fall and subsequent quenching.
\citet[][]{Bachmann+2021} showed that distant UDGs in clusters are 
relatively under-abundant, as compared to local UDGs, by a factor $\sim3$.

\begin{figure*}[!ht]
\centering
\includegraphics[scale=0.5]{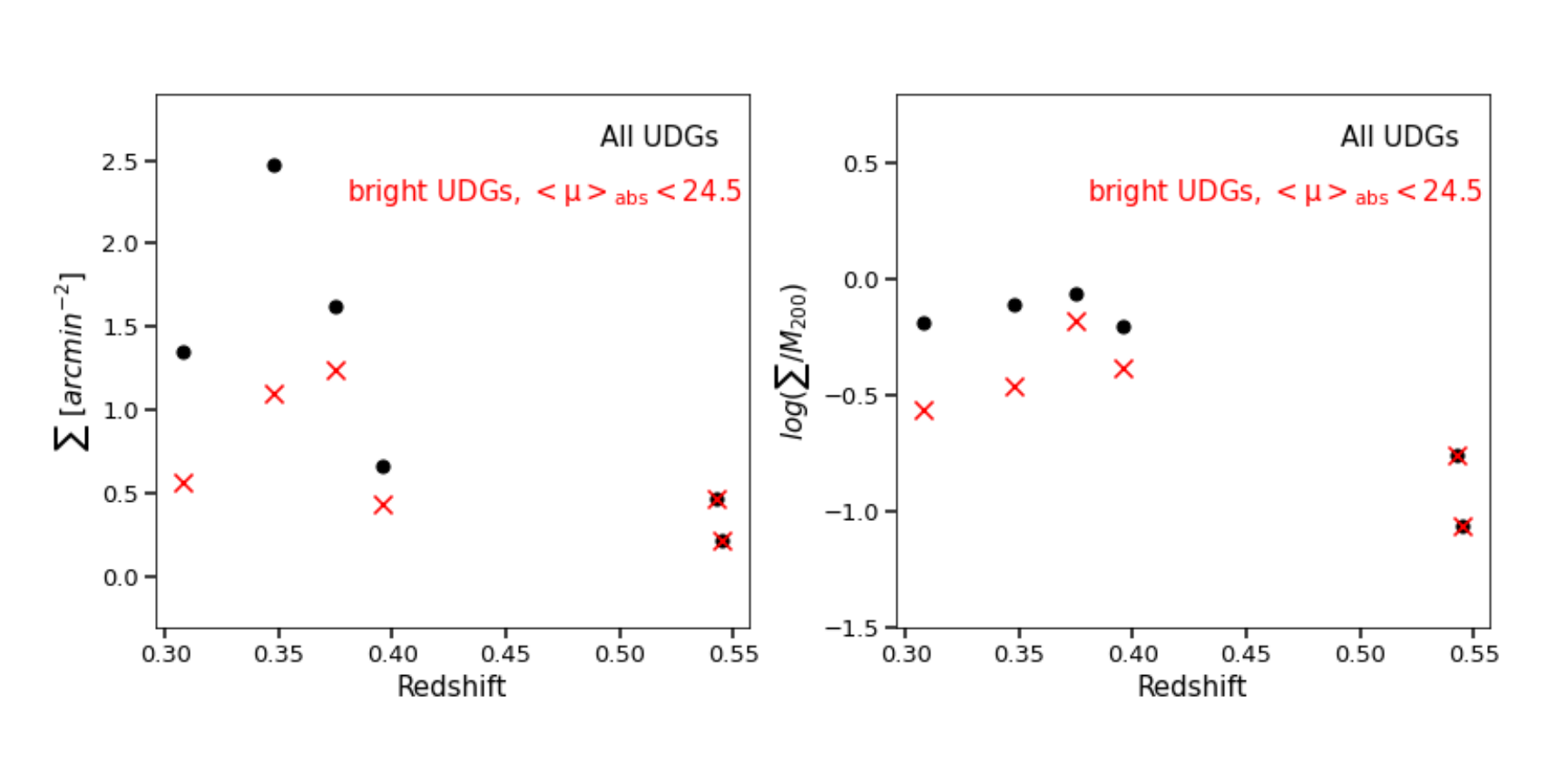}
\caption{Surface number densities of UDGs in HFF clusters as a function of redshift.
Left panel shows the completenes-corrected values and 
right panel shows the cluster-mass-calibrated $\&$ completeness-corrected values, the unit of y-axis in two panels are $arcmin^{-2}$ and log[$arcmin^{-2}*10^{-15}M_{sun}$] from left to right.
Black points and red crosses show the surface number densities computed using all UDG sample and bright UDGs ($\langle \mu \rangle_{abs} <24.5$), respectively.
}
\label{fig:Dens}
\end{figure*}

In Fig.\,\ref{fig:Dens}, we show the surface densities of UDGs in HFF clusters as a function of redshift.
The density value of each cluster is computed with the following formula: 
\begin{equation}
\sum = \frac{\sum_{n}^{N_{UDGs,clu}}1/comp_{n}}{Area_{clu}}
\end{equation}
Here, $N_{UDGs,clu}$ is the number of UDGs in each cluster. 
$comp_{n}$ is the completeness value, which is determined from Fig.\,\ref{fig:completeness} according to the surface brightness and effective radius of each UDG.
$Area_{clu}$ is the coverage area of HFF cluster field in F814W band, they are from the Table 1 of \citet[][]{Shipley+2018}.
The result of completeness-corrected surface number densities of UDGs ($\sum$) is shown in the left panel of Fig.\,\ref{fig:Dens} (black points). 
There is an obvious difference of $\sum$ between clusters at higher redshift and lower redshift. 
\citet[][]{vanderburg+2016} found that the abundance of UDGs are correlated with the virial mass of host clusters .
In the right panel of Fig.\,\ref{fig:Dens}, we calibrate $\sum$ with the M200 of each cluster.
Here, we adopt the same M200 values for HFF clusters as listed in Table 1 of \citet[][]{Janssens+2019}. 
The M200-calibrated $\sum$ of clusters at z$\sim$0.55
is smaller than those at $z<0.4$, the difference is greater than 0.55 dex (black points).
Considering object will look dimmer when it is put at a high redshift due to cosmic dimming effect, the difference of surface number densities of UDGs shown with the black points in Fig.\,\ref{fig:Dens} could be a result of systematic effect.
In order to check this, we compute surface number densities only for bright UDGs ($\langle \mu \rangle_{abs} <24.5$), the results are plotted with red cross in Fig.\,\ref{fig:Dens}. 
The limit of 24.5 we used here is close to the faitest UDG we found in two $z\sim 0.55$ clusters MACS0717 and MACS1149, above this level, our UDGs have completeness greater than 90\%,
The difference in surface number densities for bright UDGs between high-z clusters (z$\sim$0.55) and low-z clusters ($z<0.4$) still exists.
Based on \citet[][]{Rong+2017}' simulation, cluster UDGs can be from the infall of field-born UDGs, 
and the median infall time predicted in their work is $\sim$8.9\,Gyr (corresponding to $z\sim 0.43$).
The lack of UDGs in clusters MACS0717 and MACS1149 could be a result that few UDGs have fell into dense environment at $z> 0.4$, though large uncertainties exist.
Further exploration with better observations is needed in the future.


\section{Summary} \label{sec:sum}

We carefully identify 108 UDGs from six distant massive galaxy clusters
in the HFF in redshift range from 0.308 to 0.545. 
We measure their structral parameters using GALFIT and their radial rest-frame color profiles, 
and make a comparison with UDGs in the Coma cluster. 
We show that the HFF UDGs have a median S\'ersic index of 1.09, which is close to 
0.86 for Coman UDGs. The median axis-ratio value is 0.68 for HFF UDGs and 0.74 for Coma UDGs, respectively.
We find that UDGs in the HFF do not show significantly large color gradients
within their effective radii. Changes from inside to outside of the median color profile 
are smaller than 0.1 magnitudes. Meanwhile, unlike UDGs in the Coma cluster, 
whose color profiles are mostly red from inside to outside, 
a large fraction of HFF UDGs have blue colors and are star-forming. 
Our findings provide evidence that UDGs in clusters may have a
self-similar star formation quenching mode when evolving from distant 
to the local universe.
Besides, we find the M200-calibrated surface number densities of UDGs is lower at two $z\sim 0.55$ clusters when comparing to other HFF clusters. 
Under the scenario that UDGs might be born in the field and finnally infall into galaxy clusters \citep[][]{Roman+2017},
the lack of UDGs found in distant clusters imply that few UDGs have fell into dense environment at $z>0.4$,
which agrees with the simulation work from \citet[][]{Rong+2017}.

%


\startlongtable
\begin{deluxetable*}{llllllll}
\tabletypesize{\scriptsize}
\tablecaption{Catalog of 108 UDGs identified in the HFF program \label{tab:UDGcat}}
\tablecolumns{8}
\tablenum{3}
\tablewidth{50pt}
\tablehead{
\colhead{ID} &
\colhead{$R.A.(J2000)$} &
\colhead{$Dec.(J2000)$} &
\colhead{$<\mu>$} &
\colhead{$R_{e,SMA}$} & \colhead{n} & \colhead{q} & \colhead{rest-frame U-V}
}
\startdata
A2744clu0112 & 3.5817 & -30.4312 & 24.28 & 2.0 & 1.1 & 0.4 & 0.85\\
A2744clu0448 & 3.5952 & -30.4228 & 24.88 & 1.64 & 0.95 & 0.79 & 0.56\\
A2744clu0679 & 3.5912 & -30.4194 & 24.91 & 1.94 & 1.31 & 0.58 & 2.14\\
A2744clu0745 & 3.5787 & -30.4188 & 24.75 & 2.71 & 1.72 & 0.58 & 1.22\\
A2744clu1656 & 3.6062 & -30.4116 & 24.11 & 2.43 & 0.71 & 0.34 & 0.69\\
A2744clu1717 & 3.616 & -30.4107 & 24.98 & 2.44 & 1.58 & 0.64 & 0.91\\
A2744clu2029 & 3.5748 & -30.4095 & 24.83 & 2.22 & 1.36 & 0.55 & 1.36\\
A2744clu2489 & 3.5596 & -30.4065 & 25.13 & 2.1 & 0.75 & 0.54 & 1.39\\
A2744clu2651 & 3.6231 & -30.406 & 24.54 & 1.9 & 0.79 & 0.72 & 0.64\\
A2744clu4431 & 3.5648 & -30.397 & 24.99 & 1.69 & 1.05 & 0.97 & 1.17\\
A2744clu5026 & 3.5641 & -30.3944 & 24.65 & 1.82 & 0.79 & 0.78 & 0.59\\
A2744clu5159 & 3.5638 & -30.3935 & 24.99 & 1.75 & 0.79 & 0.72 & 1.33\\
A2744clu6355 & 3.6024 & -30.387 & 24.16 & 1.81 & 3.0 & 0.89 & 1.1\\
A2744clu6625 & 3.5822 & -30.3849 & 24.37 & 3.28 & 1.93 & 0.34 & 1.05\\
A2744clu7053 & 3.5634 & -30.3815 & 25.17 & 1.56 & 0.79 & 0.87 & 0.66\\
A2744clu7089 & 3.5835 & -30.3822 & 24.63 & 2.55 & 0.85 & 0.49 & 1.37\\
A2744clu7219 & 3.5665 & -30.3806 & 24.11 & 1.72 & 1.29 & 0.8 & 0.52\\
A2744clu7257 & 3.5641 & -30.3817 & 24.03 & 3.15 & 1.23 & 0.76 & 1.06\\
A2744clu7651 & 3.5914 & -30.3774 & 24.53 & 1.8 & 1.04 & 0.86 & 1.31\\
A2744clu7696 & 3.5907 & -30.3772 & 24.63 & 3.62 & 0.84 & 0.55 & 0.68\\
A2744clu8134 & 3.5762 & -30.3718 & 24.18 & 1.69 & 0.75 & 0.7 & 0.87\\
A2744clu8312 & 3.6025 & -30.3695 & 24.15 & 1.84 & 1.31 & 0.72 & 0.6\\
A2744clu8655 & 3.5908 & -30.3638 & 24.27 & 1.78 & 1.73 & 0.43 & 1.21\\
A2744clu8657 & 3.595 & -30.3638 & 24.49 & 1.62 & 1.81 & 0.79 & 0.86\\
A2744clu8681 & 3.5853 & -30.3644 & 24.22 & 3.22 & 1.55 & 0.43 & 0.9\\
A2744clu8818 & 3.5899 & -30.3622 & 24.34 & 3.24 & 1.08 & 0.9 & 1.04\\
A370clu0353 & 39.9598 & -1.6067 & 24.81 & 2.78 & 0.59 & 0.91 & 0.27\\
A370clu0459 & 39.9636 & -1.6043 & 24.1 & 1.84 & 0.82 & 0.51 & 1.08\\
A370clu0646 & 39.9604 & -1.6014 & 24.11 & 2.18 & 1.62 & 0.52 & 0.74\\
A370clu0896 & 39.9838 & -1.5979 & 25.2 & 4.19 & 1.94 & 0.6 & 0.75\\
A370clu1046 & 39.9821 & -1.5958 & 24.16 & 2.64 & 1.67 & 0.4 & 1.26\\
A370clu1456 & 39.9851 & -1.5917 & 24.12 & 2.3 & 1.14 & 0.52 & 0.67\\
A370clu1760 & 39.9958 & -1.5885 & 24.15 & 1.7 & 1.23 & 0.68 & 1.53\\
A370clu2123 & 39.9822 & -1.5854 & 25.25 & 3.04 & 1.11 & 0.61 & 1.02\\
A370clu2258 & 39.9887 & -1.5837 & 24.93 & 3.03 & 0.65 & 0.67 & 1.65\\
A370clu2416 & 39.9451 & -1.5827 & 24.38 & 2.11 & 1.29 & 0.63 & 1.48\\
A370clu2512 & 39.9897 & -1.582 & 24.17 & 1.69 & 2.08 & 0.59 & 1.65\\
A370clu2569 & 39.9863 & -1.582 & 24.07 & 2.09 & 1.33 & 0.84 & 1.01\\
A370clu3299 & 39.9402 & -1.5762 & 24.51 & 1.82 & 0.94 & 0.89 & 0.87\\
A370clu3386 & 39.9506 & -1.5754 & 24.13 & 1.66 & 0.69 & 0.51 & 0.75\\
A370clu3876 & 39.9615 & -1.573 & 24.28 & 2.01 & 0.73 & 0.91 & 0.9\\
A370clu3936 & 39.941 & -1.5714 & 24.04 & 1.65 & 1.12 & 0.75 & 0.83\\
A370clu3999 & 39.9523 & -1.5708 & 24.27 & 1.72 & 0.95 & 0.56 & 1.15\\
A370clu4169 & 39.9545 & -1.5696 & 24.28 & 1.97 & 2.07 & 0.7 & 0.87\\
A370clu4746 & 39.9717 & -1.5646 & 24.44 & 1.7 & 2.14 & 0.89 & 0.82\\
A370clu4938 & 39.9344 & -1.5626 & 24.2 & 2.32 & 0.62 & 0.56 & 0.76\\
A370clu5038 & 39.9491 & -1.5621 & 24.45 & 3.58 & 1.36 & 0.39 & 1.25\\
A370clu5094 & 39.9827 & -1.5611 & 24.07 & 1.55 & 1.84 & 0.58 & 0.6\\
A370clu5325 & 39.9841 & -1.559 & 24.14 & 1.75 & 1.41 & 0.86 & 1.26\\
AS1063clu0008 & 342.178 & -44.5698 & 24.89 & 2.23 & 1.92 & 0.88 & 0.89\\
AS1063clu0224 & 342.1824 & -44.5616 & 25.29 & 2.7 & 1.86 & 0.74 & 0.71\\
AS1063clu0288 & 342.1791 & -44.5603 & 24.05 & 1.65 & 1.83 & 0.72 & 1.15\\
AS1063clu0308 & 342.175 & -44.56 & 24.34 & 1.71 & 0.8 & 0.42 & 0.26\\
AS1063clu0379 & 342.1697 & -44.5584 & 24.03 & 1.88 & 1.1 & 0.43 & 1.11\\
AS1063clu0496 & 342.1641 & -44.5562 & 24.19 & 1.79 & 0.88 & 0.26 & 0.71\\
AS1063clu1208 & 342.17 & -44.5469 & 24.53 & 1.86 & 1.55 & 0.69 & 0.59\\
AS1063clu1228 & 342.1644 & -44.5464 & 24.16 & 1.64 & 1.3 & 0.5 & 1.07\\
AS1063clu2393 & 342.199 & -44.5381 & 24.25 & 2.19 & 0.93 & 0.52 & 1.3\\
AS1063clu2427 & 342.1948 & -44.538 & 24.22 & 1.93 & 0.73 & 0.76 & 1.06\\
AS1063clu2749 & 342.2347 & -44.5355 & 24.81 & 1.93 & 0.82 & 0.57 & 1.16\\
AS1063clu2812 & 342.1494 & -44.5352 & 24.1 & 1.73 & 2.0 & 0.91 & 1.56\\
AS1063clu2960 & 342.203 & -44.5349 & 24.67 & 1.57 & 0.78 & 0.95 & 0.43\\
AS1063clu3056 & 342.1991 & -44.5344 & 24.63 & 2.05 & 0.71 & 0.59 & 0.79\\
AS1063clu3122 & 342.1441 & -44.5336 & 24.79 & 2.42 & 1.28 & 0.78 & 1.04\\
AS1063clu3242 & 342.2192 & -44.5329 & 24.25 & 1.88 & 0.35 & 0.45 & 0.81\\
AS1063clu3267 & 342.1437 & -44.5332 & 24.2 & 3.21 & 0.87 & 0.46 & 0.73\\
AS1063clu3377 & 342.2192 & -44.5322 & 24.82 & 2.04 & 0.87 & 0.7 & 1.49\\
AS1063clu3447 & 342.2319 & -44.532 & 24.25 & 2.14 & 1.17 & 0.6 & 1.43\\
AS1063clu3471 & 342.2271 & -44.5318 & 24.65 & 2.27 & 0.97 & 0.76 & 1.1\\
AS1063clu3607 & 342.217 & -44.531 & 25.18 & 2.44 & 0.56 & 0.44 & -0.45\\
AS1063clu3937 & 342.2027 & -44.5308 & 24.73 & 2.57 & 1.75 & 0.79 & 2.07\\
AS1063clu4009 & 342.1379 & -44.5295 & 24.12 & 1.57 & 1.01 & 0.7 & 0.74\\
AS1063clu4519 & 342.2163 & -44.5273 & 24.02 & 1.59 & 0.96 & 0.94 & 1.27\\
AS1063clu4855 & 342.1356 & -44.5254 & 24.16 & 2.16 & 1.34 & 0.81 & 0.72\\
AS1063clu4972 & 342.1483 & -44.5243 & 24.52 & 1.73 & 1.1 & 0.6 & 1.12\\
AS1063clu5030 & 342.1747 & -44.5249 & 24.8 & 1.78 & 1.54 & 0.92 & 0.84\\
AS1063clu5710 & 342.1502 & -44.5194 & 24.56 & 2.21 & 0.87 & 0.76 & 0.73\\
AS1063clu5943 & 342.1656 & -44.5179 & 24.78 & 2.61 & 2.57 & 0.7 & 1.04\\
AS1063clu6074 & 342.1734 & -44.5171 & 24.7 & 2.76 & 2.9 & 0.76 & 1.1\\
AS1063clu6396 & 342.1882 & -44.5143 & 24.08 & 2.03 & 0.75 & 0.91 & 1.1\\
AS1063clu6652 & 342.1866 & -44.5171 & 24.08 & 1.97 & 2.27 & 0.84 & 1.45\\
AS1063clu6653 & 342.1972 & -44.5108 & 24.81 & 2.18 & 0.47 & 0.63 & 1.66\\
AS1063clu6721 & 342.1967 & -44.5102 & 24.82 & 1.66 & 0.69 & 0.75 & 0.6\\
AS1063clu6800 & 342.1955 & -44.5095 & 24.04 & 2.23 & 0.7 & 0.94 & 0.62\\
AS1063clu7141 & 342.1833 & -44.5037 & 24.9 & 2.86 & 0.75 & 0.69 & 1.32\\
M0416clu0656 & 64.0465 & -24.095 & 24.01 & 1.56 & 0.87 & 0.62 & 0.92\\
M0416clu0666 & 64.045 & -24.0954 & 24.24 & 2.27 & 2.05 & 0.57 & 0.9\\
M0416clu0833 & 64.0661 & -24.0925 & 24.71 & 1.51 & 0.57 & 0.75 & 1.13\\
M0416clu1090 & 64.0126 & -24.0899 & 24.04 & 1.66 & 1.04 & 0.64 & 0.89\\
M0416clu4132 & 64.0078 & -24.0707 & 24.32 & 1.52 & 2.08 & 0.71 & 1.01\\
M0416clu5295 & 64.0575 & -24.0643 & 24.1 & 1.97 & 0.69 & 0.66 & 1.21\\
M0416clu5483 & 64.0241 & -24.0627 & 24.01 & 1.62 & 0.96 & 0.84 & 1.11\\
M0416clu6651 & 64.0572 & -24.0532 & 24.65 & 2.24 & 0.3 & 0.54 & 1.2\\
M0416clu6894 & 64.0427 & -24.0504 & 24.91 & 3.71 & 0.82 & 0.64 & 1.07\\
M0717clu0069 & 109.4029 & 37.7197 & 24.13 & 2.51 & 1.76 & 0.44 & 1.22\\
M0717clu0456 & 109.42 & 37.7254 & 24.01 & 1.68 & 1.0 & 0.62 & 1.78\\
M0717clu1415 & 109.3851 & 37.7338 & 24.06 & 3.09 & 1.15 & 0.98 & 1.46\\
M0717clu5158 & 109.3811 & 37.7647 & 24.19 & 1.75 & 2.4 & 0.65 & 0.66\\
M0717clu5661 & 109.3872 & 37.7699 & 24.16 & 2.85 & 1.2 & 0.91 & 1.19\\
M0717clu5958 & 109.3839 & 37.7733 & 24.27 & 2.48 & 0.83 & 0.77 & 0.66\\
M1149clu0324 & 177.4102 & 22.3731 & 24.09 & 2.4 & 2.02 & 0.85 & 0.71\\
M1149clu0541 & 177.4016 & 22.3764 & 24.43 & 2.63 & 0.35 & 0.48 & 1.0\\
M1149clu0778 & 177.3933 & 22.3794 & 24.02 & 1.91 & 2.08 & 0.88 & 2.28\\
M1149clu3274 & 177.4119 & 22.398 & 24.15 & 2.76 & 1.03 & 0.62 & 1.31\\
M1149clu3831 & 177.3808 & 22.4017 & 24.46 & 1.89 & 2.38 & 0.65 & 0.52\\
M1149clu5184 & 177.4058 & 22.4106 & 24.29 & 2.15 & 0.87 & 0.66 & 0.7\\
M1149clu5625 & 177.3822 & 22.4142 & 24.04 & 2.4 & 1.89 & 0.39 & 1.09\\
M1149clu6156 & 177.4072 & 22.4199 & 24.06 & 2.56 & 2.69 & 0.47 & 1.2\\
\enddata
\tablecomments{Basic information of our 108 UDGs.
Col `ID' is the combined ID of cluster name and id from Shipley's catalog.
R.A.(J2000) and Dec.(J2000) are directly from Shipley's catalog.
$<\mu>$, $R_{e,SMA}$, n, q are structural parameters.
}
\end{deluxetable*}


\acknowledgments

We thank anonymous referee for the insightful suggestions, 
which significantly helped us improve this paper.
We thank Xianmin Meng, Xin Zhang and Juanjuan Ren for useful suggestions and discussions.
This project is supported by the National Natural Science Foundation of China 
(NSFC grants Nos.12273052,11733006, U1931109, 12090041, 12090040), 
the National Key R\&D Program of China (No. 2017YFA0402704), 
and the science research grants from the China Manned Space Project
(NOs. CMS-CSST-2021-A04, CMS-CSST-2021-B06).
This study is based on observations obtained with the NASA/ESA Hubble Space Telescope, retrieved from the Mikulski Archive for Space Telescopes (MAST) at the Space Telescope Science Institute (STScI). 
This work is based on data and catalog products from HFF-DeepSpace, funded by the National Science Foundation and Space Telescope Science Institute.
STScI is operated by the Association of Universities for Research in Astronomy, Inc. under NASA contract NAS 5-26555.

\clearpage

\bibliographystyle{aasjournal}
\bibliography{journals}

\begin{thebibliography}{}
\expandafter\ifx\csname natexlab\endcsname\relax\def\natexlab#1{#1}\fi
\providecommand{\url}[1]{\href{#1}{#1}}

\bibitem[{{Akhlaghi} \& {Ichikawa}(2015)}]{Akhlaghi+2015}
{Akhlaghi}, M., \& {Ichikawa}, T. 2015, \apjs, 220, 1

\bibitem[{{Amorisco} {et~al.}(2018){Amorisco}, {Monachesi}, {Agnello}, \&
  {White}}]{Amorisco2018}
{Amorisco}, N.~C., {Monachesi}, A., {Agnello}, A., \& {White}, S.~D.~M. 2018,
  \mnras, 475, 4235

\bibitem[{{Bachmann} {et~al.}(2021){Bachmann}, {van der Burg}, {Fensch},
  {Brammer}, \& {Muzzin}}]{Bachmann+2021}
{Bachmann}, A., {van der Burg}, R. F.~J., {Fensch}, J., {Brammer}, G., \&
  {Muzzin}, A. 2021, \aap, 646, L12

\bibitem[{{Barden} {et~al.}(2012){Barden}, {H{\"a}u{\ss}ler}, {Peng},
  {McIntosh}, \& {Guo}}]{Barden+2012}
{Barden}, M., {H{\"a}u{\ss}ler}, B., {Peng}, C.~Y., {McIntosh}, D.~H., \&
  {Guo}, Y. 2012, {GALAPAGOS: Galaxy Analysis over Large Areas: Parameter
  Assessment by GALFITting Objects from SExtractor}, Astrophysics Source Code
  Library, record ascl:1203.002, , , ascl:1203.002

\bibitem[{{Bertin} \& {Arnouts}(1996)}]{BertinArnouts1996}
{Bertin}, E., \& {Arnouts}, S. 1996, \aaps, 117, 393

\bibitem[{{Blanton} {et~al.}(2003){Blanton}, {Brinkmann}, {Csabai}, {Doi},
  {Eisenstein}, {Fukugita}, {Gunn}, {Hogg}, \& {Schlegel}}]{blanton+2003}
{Blanton}, M.~R., {Brinkmann}, J., {Csabai}, I., {et~al.} 2003, \aj, 125, 2348

\bibitem[{{Brammer} {et~al.}(2008){Brammer}, {van Dokkum}, \&
  {Coppi}}]{Brammer+2008}
{Brammer}, G.~B., {van Dokkum}, P.~G., \& {Coppi}, P. 2008, \apj, 686, 1503

\bibitem[{{Carter} {et~al.}(2008){Carter}, {Goudfrooij}, {Mobasher},
  {Ferguson}, {Puzia}, {Aguerri}, {Balcells}, {Batcheldor}, {Bridges},
  {Davies}, {Erwin}, {Graham}, {Guzm{\'a}n}, {Hammer}, {Hornschemeier},
  {Hoyos}, {Hudson}, {Huxor}, {Jogee}, {Komiyama}, {Lotz}, {Lucey}, {Marzke},
  {Merritt}, {Miller}, {Miller}, {Mouhcine}, {Okamura}, {Peletier},
  {Phillipps}, {Poggianti}, {Sharples}, {Smith}, {Trentham}, {Tully},
  {Valentijn}, \& {Verdoes Kleijn}}]{Carter+2008}
{Carter}, D., {Goudfrooij}, P., {Mobasher}, B., {et~al.} 2008, \apjs, 176, 424

\bibitem[{{Castellano} {et~al.}(2016){Castellano}, {Amor{\'\i}n}, {Merlin},
  {Fontana}, {McLure}, {M{\'a}rmol-Queralt{\'o}}, {Mortlock}, {Parsa},
  {Dunlop}, {Elbaz}, {Balestra}, {Boucaud}, {Bourne}, {Boutsia}, {Brammer},
  {Bruce}, {Buitrago}, {Capak}, {Cappelluti}, {Ciesla}, {Comastri}, {Cullen},
  {Derriere}, {Faber}, {Giallongo}, {Grazian}, {Grillo}, {Mercurio},
  {Micha{\l}owski}, {Nonino}, {Paris}, {Pentericci}, {Pilo}, {Rosati},
  {Santini}, {Schreiber}, {Shu}, \& {Wang}}]{Castellano+2016}
{Castellano}, M., {Amor{\'\i}n}, R., {Merlin}, E., {et~al.} 2016, \aap, 590,
  A31

\bibitem[{{Chilingarian}(2009)}]{Chilingarian+2009}
{Chilingarian}, I.~V. 2009, \mnras, 394, 1229

\bibitem[{{Conselice} {et~al.}(2002){Conselice}, {Gallagher}, \&
  {Wyse}}]{Conselice+2002}
{Conselice}, C.~J., {Gallagher}, John~S., I., \& {Wyse}, R. F.~G. 2002, \aj,
  123, 2246

\bibitem[{{Conselice} {et~al.}(2003){Conselice}, {Gallagher}, \&
  {Wyse}}]{Conselice+2003}
---. 2003, \aj, 125, 66

\bibitem[{{de Rijcke} {et~al.}(2009){de Rijcke}, {Penny}, {Conselice},
  {Valcke}, \& {Held}}]{deRijcke+2009}
{de Rijcke}, S., {Penny}, S.~J., {Conselice}, C.~J., {Valcke}, S., \& {Held},
  E.~V. 2009, \mnras, 393, 798

\bibitem[{{Gu} {et~al.}(2018){Gu}, {Conroy}, {Law}, {van Dokkum}, {Yan},
  {Wake}, {Bundy}, {Merritt}, {Abraham}, {Zhang}, {Bershady}, {Bizyaev},
  {Brinkmann}, {Drory}, {Grabowski}, {Masters}, {Pan}, {Parejko}, {Weijmans},
  \& {Zhang}}]{Gu+2018}
{Gu}, M., {Conroy}, C., {Law}, D., {et~al.} 2018, \apj, 859, 37

\bibitem[{{Haigh} {et~al.}(2021){Haigh}, {Chamba}, {Venhola}, {Peletier},
  {Doorenbos}, {Watkins}, \& {Wilkinson}}]{Haigh+2021}
{Haigh}, C., {Chamba}, N., {Venhola}, A., {et~al.} 2021, \aap, 645, A107

\bibitem[{{Hammer} {et~al.}(2010){Hammer}, {Verdoes Kleijn}, {Hoyos}, {den
  Brok}, {Balcells}, {Ferguson}, {Goudfrooij}, {Carter}, {Guzm{\'a}n},
  {Peletier}, {Smith}, {Graham}, {Trentham}, {Peng}, {Puzia}, {Lucey}, {Jogee},
  {Aguerri}, {Batcheldor}, {Bridges}, {Chiboucas}, {Davies}, {del Burgo},
  {Erwin}, {Hornschemeier}, {Hudson}, {Huxor}, {Jenkins}, {Karick},
  {Khosroshahi}, {Kourkchi}, {Komiyama}, {Lotz}, {Marzke}, {Marinova},
  {Matkovic}, {Merritt}, {Miller}, {Miller}, {Mobasher}, {Mouhcine}, {Okamura},
  {Percival}, {Phillipps}, {Poggianti}, {Price}, {Sharples}, {Tully}, \&
  {Valentijn}}]{Hammer+2010}
{Hammer}, D., {Verdoes Kleijn}, G., {Hoyos}, C., {et~al.} 2010, \apjs, 191, 143

\bibitem[{{He} {et~al.}(2019){He}, {Wu}, {Du}, {Wicker}, {Zhao}, {Lei}, \&
  {Liu}}]{He+2019}
{He}, M., {Wu}, H., {Du}, W., {et~al.} 2019, \apj, 880, 30

\bibitem[{{Hogg} {et~al.}(2002){Hogg}, {Baldry}, {Blanton}, \&
  {Eisenstein}}]{hogg+2002}
{Hogg}, D.~W., {Baldry}, I.~K., {Blanton}, M.~R., \& {Eisenstein}, D.~J. 2002,
  arXiv e-prints, astro

\bibitem[{{Iodice} {et~al.}(2020){Iodice}, {Cantiello}, {Hilker}, {Rejkuba},
  {Arnaboldi}, {Spavone}, {Greggio}, {Forbes}, {D'Ago}, {Mieske}, {Spiniello},
  {La Marca}, {Rampazzo}, {Paolillo}, {Capaccioli}, \&
  {Schipani}}]{Iodice+2020}
{Iodice}, E., {Cantiello}, M., {Hilker}, M., {et~al.} 2020, \aap, 642, A48

\bibitem[{{Janssens} {et~al.}(2017){Janssens}, {Abraham}, {Brodie}, {Forbes},
  {Romanowsky}, \& {van Dokkum}}]{Janssens+2017}
{Janssens}, S., {Abraham}, R., {Brodie}, J., {et~al.} 2017, \apjl, 839, L17

\bibitem[{{Janssens} {et~al.}(2019){Janssens}, {Abraham}, {Brodie}, {Forbes},
  \& {Romanowsky}}]{Janssens+2019}
{Janssens}, S.~R., {Abraham}, R., {Brodie}, J., {Forbes}, D.~A., \&
  {Romanowsky}, A.~J. 2019, \apj, 887, 92

\bibitem[{{Jerjen} {et~al.}(2000){Jerjen}, {Binggeli}, \&
  {Freeman}}]{Jerjen+2000}
{Jerjen}, H., {Binggeli}, B., \& {Freeman}, K.~C. 2000, \aj, 119, 593

\bibitem[{{Kadowaki} {et~al.}(2017){Kadowaki}, {Zaritsky}, \&
  {Donnerstein}}]{Kadowaki+2017}
{Kadowaki}, J., {Zaritsky}, D., \& {Donnerstein}, R.~L. 2017, \apjl, 838, L21

\bibitem[{{Kadowaki} {et~al.}(2021){Kadowaki}, {Zaritsky}, {Donnerstein}, {RS},
  {Karunakaran}, \& {Spekkens}}]{Kadowaki+2021}
{Kadowaki}, J., {Zaritsky}, D., {Donnerstein}, R.~L., {et~al.} 2021, \apj, 923,
  257

\bibitem[{{Koda} {et~al.}(2015){Koda}, {Yagi}, {Yamanoi}, \&
  {Komiyama}}]{Koda+15}
{Koda}, J., {Yagi}, M., {Yamanoi}, H., \& {Komiyama}, Y. 2015, \apjl, 807, L2

\bibitem[{{Koleva} {et~al.}(2011){Koleva}, {Prugniel}, {De Rijcke}, \&
  {Zeilinger}}]{Koleva+2011}
{Koleva}, M., {Prugniel}, P., {De Rijcke}, S., \& {Zeilinger}, W.~W. 2011,
  \mnras, 417, 1643

\bibitem[{{Lee} {et~al.}(2020){Lee}, {Kang}, {Lee}, \& {Jang}}]{Lee+2020}
{Lee}, J.~H., {Kang}, J., {Lee}, M.~G., \& {Jang}, I.~S. 2020, \apj, 894, 75

\bibitem[{{Lee} {et~al.}(2017){Lee}, {Kang}, {Lee}, \& {Jang}}]{Lee+2017}
{Lee}, M.~G., {Kang}, J., {Lee}, J.~H., \& {Jang}, I.~S. 2017, \apj, 844, 157

\bibitem[{{Leisman} {et~al.}(2017){Leisman}, {Haynes}, {Janowiecki},
  {Hallenbeck}, {J{\'o}zsa}, {Giovanelli}, {Adams}, {Bernal Neira}, {Cannon},
  {Janesh}, {Rhode}, \& {Salzer}}]{Leisman+2017}
{Leisman}, L., {Haynes}, M.~P., {Janowiecki}, S., {et~al.} 2017, \apj, 842, 133

\bibitem[{{Liu} {et~al.}(2016){Liu}, {Jiang}, {Guo}, {Koo}, {Faber}, {Zheng},
  {Yesuf}, {Barro}, {Li}, {Li}, {Wang}, {Mao}, \& {Fang}}]{Liu+2016}
{Liu}, F.~S., {Jiang}, D., {Guo}, Y., {et~al.} 2016, \apjl, 822, L25

\bibitem[{{Liu} {et~al.}(2017){Liu}, {Jiang}, {Faber}, {Koo}, {Yesuf},
  {Tacchella}, {Mao}, {Wang}, {Guo}, {Fang}, {Barro}, {Zheng}, {Jia}, {Tong},
  {Liu}, \& {Meng}}]{Liu+2017}
{Liu}, F.~S., {Jiang}, D., {Faber}, S.~M., {et~al.} 2017, \apjl, 844, L2

\bibitem[{{Liu} {et~al.}(2018){Liu}, {Jia}, {Yesuf}, {Faber}, {Koo}, {Guo},
  {Bell}, {Jiang}, {Wang}, {Koekemoer}, {Zheng}, {Fang}, {Barro},
  {P{\'e}rez-Gonz{\'a}lez}, {Dekel}, {Kocevski}, {Hathi}, {Croton},
  {Huertas-Company}, {Meng}, {Tong}, \& {Liu}}]{Liu+2018}
{Liu}, F.~S., {Jia}, M., {Yesuf}, H.~M., {et~al.} 2018, \apj, 860, 60

\bibitem[{{Lotz} {et~al.}(2017){Lotz}, {Koekemoer}, {Coe}, {Grogin}, {Capak},
  {Mack}, {Anderson}, {Avila}, {Barker}, {Borncamp}, {Brammer}, {Durbin},
  {Gunning}, {Hilbert}, {Jenkner}, {Khandrika}, {Levay}, {Lucas}, {MacKenty},
  {Ogaz}, {Porterfield}, {Reid}, {Robberto}, {Royle}, {Smith},
  {Storrie-Lombardi}, {Sunnquist}, {Surace}, {Taylor}, {Williams}, {Bullock},
  {Dickinson}, {Finkelstein}, {Natarajan}, {Richard}, {Robertson}, {Tumlinson},
  {Zitrin}, {Flanagan}, {Sembach}, {Soifer}, \& {Mountain}}]{Lotz+2017}
{Lotz}, J.~M., {Koekemoer}, A., {Coe}, D., {et~al.} 2017, \apj, 837, 97

\bibitem[{{Merlin} {et~al.}(2016){Merlin}, {Amor{\'\i}n}, {Castellano},
  {Fontana}, {Buitrago}, {Dunlop}, {Elbaz}, {Boucaud}, {Bourne}, {Boutsia},
  {Brammer}, {Bruce}, {Capak}, {Cappelluti}, {Ciesla}, {Comastri}, {Cullen},
  {Derriere}, {Faber}, {Ferguson}, {Giallongo}, {Grazian}, {Lotz},
  {Micha{\l}owski}, {Paris}, {Pentericci}, {Pilo}, {Santini}, {Schreiber},
  {Shu}, \& {Wang}}]{Merlin+2016}
{Merlin}, E., {Amor{\'\i}n}, R., {Castellano}, M., {et~al.} 2016, \aap, 590,
  A30

\bibitem[{{Mieske} {et~al.}(2007){Mieske}, {Hilker}, {Infante}, \& {Mendes de
  Oliveira}}]{Mieske+2007}
{Mieske}, S., {Hilker}, M., {Infante}, L., \& {Mendes de Oliveira}, C. 2007,
  \aap, 463, 503

\bibitem[{{Mihos} {et~al.}(2015){Mihos}, {Durrell}, {Ferrarese}, {Feldmeier},
  {C{\^o}t{\'e}}, {Peng}, {Harding}, {Liu}, {Gwyn}, \&
  {Cuillandre}}]{Mihos+2015}
{Mihos}, J.~C., {Durrell}, P.~R., {Ferrarese}, L., {et~al.} 2015, \apjl, 809,
  L21

\bibitem[{{Pagul} {et~al.}(2021){Pagul}, {S{\'a}nchez}, {Davidzon}, \&
  {Mobasher}}]{Pagul+2021}
{Pagul}, A., {S{\'a}nchez}, F.~J., {Davidzon}, I., \& {Mobasher}, B. 2021,
  \apjs, 256, 27

\bibitem[{{Peng} {et~al.}(2002){Peng}, {Ho}, {Impey}, \& {Rix}}]{Peng+2002}
{Peng}, C.~Y., {Ho}, L.~C., {Impey}, C.~D., \& {Rix}, H.-W. 2002, \aj, 124, 266

\bibitem[{{Peng} {et~al.}(2010){Peng}, {Ho}, {Impey}, \& {Rix}}]{Peng+2010}
---. 2010, \aj, 139, 2097

\bibitem[{{Penny} {et~al.}(2009){Penny}, {Conselice}, {de Rijcke}, \&
  {Held}}]{Penny+2009}
{Penny}, S.~J., {Conselice}, C.~J., {de Rijcke}, S., \& {Held}, E.~V. 2009,
  \mnras, 393, 1054

\bibitem[{{Penny} {et~al.}(2011){Penny}, {Conselice}, {de Rijcke}, {Held},
  {Gallagher}, \& {O'Connell}}]{Penny+2011}
{Penny}, S.~J., {Conselice}, C.~J., {de Rijcke}, S., {et~al.} 2011, \mnras,
  410, 1076

\bibitem[{{Rom{\'a}n} \& {Trujillo}(2017)}]{Roman+2017}
{Rom{\'a}n}, J., \& {Trujillo}, I. 2017, \mnras, 468, 4039

\bibitem[{{Rong} {et~al.}(2017){Rong}, {Guo}, {Gao}, {Liao}, {Xie}, {Puzia},
  {Sun}, \& {Pan}}]{Rong+2017}
{Rong}, Y., {Guo}, Q., {Gao}, L., {et~al.} 2017, \mnras, 470, 4231

\bibitem[{{Rong} {et~al.}(2020){Rong}, {Zhu}, {Johnston}, {Zhang}, {Cao},
  {Puzia}, \& {Galaz}}]{Rong+2020}
{Rong}, Y., {Zhu}, K., {Johnston}, E.~J., {et~al.} 2020, \apjl, 899, L12

\bibitem[{{Sandage} \& {Binggeli}(1984)}]{SandageBinggeli84}
{Sandage}, A., \& {Binggeli}, B. 1984, AJ, 89, 919

\bibitem[{{Santini} {et~al.}(2015){Santini}, {Ferguson}, {Fontana}, {Mobasher},
  {Barro}, {Castellano}, {Finkelstein}, {Grazian}, {Hsu}, {Lee}, {Lee},
  {Pforr}, {Salvato}, {Wiklind}, {Wuyts}, {Almaini}, {Cooper}, {Galametz},
  {Weiner}, {Amorin}, {Boutsia}, {Conselice}, {Dahlen}, {Dickinson},
  {Giavalisco}, {Grogin}, {Guo}, {Hathi}, {Kocevski}, {Koekemoer},
  {Kurczynski}, {Merlin}, {Mortlock}, {Newman}, {Paris}, {Pentericci},
  {Simons}, \& {Willner}}]{Santini+15}
{Santini}, P., {Ferguson}, H.~C., {Fontana}, A., {et~al.} 2015, \apj, 801, 97

\bibitem[{{S{\'e}rsic}(1968)}]{Sersic1968}
{S{\'e}rsic}, J.~L. 1968, {Atlas de Galaxias Australes}

\bibitem[{{Shi} {et~al.}(2017){Shi}, {Zheng}, {Zhao}, {Pan}, {Li}, {Zou},
  {Zhou}, {Guo}, {An}, \& {Li}}]{Shi+2017}
{Shi}, D.~D., {Zheng}, X.~Z., {Zhao}, H.~B., {et~al.} 2017, \apj, 846, 26

\bibitem[{{Shipley} {et~al.}(2018){Shipley}, {Lange-Vagle}, {Marchesini},
  {Brammer}, {Ferrarese}, {Stefanon}, {Kado-Fong}, {Whitaker}, {Oesch},
  {Feinstein}, {Labb{\'e}}, {Lundgren}, {Martis}, {Muzzin}, {Nedkova},
  {Skelton}, \& {van der Wel}}]{Shipley+2018}
{Shipley}, H.~V., {Lange-Vagle}, D., {Marchesini}, D., {et~al.} 2018, \apjs,
  235, 14

\bibitem[{{Thompson} \& {Gregory}(1993)}]{Thompson+1993}
{Thompson}, L.~A., \& {Gregory}, S.~A. 1993, \aj, 106, 2197

\bibitem[{{Tremmel} {et~al.}(2020){Tremmel}, {Wright}, {Brooks}, {Munshi},
  {Nagai}, \& {Quinn}}]{Tremmel20}
{Tremmel}, M., {Wright}, A.~C., {Brooks}, A.~M., {et~al.} 2020, \mnras, 497,
  2786

\bibitem[{{Trujillo} {et~al.}(2017){Trujillo}, {Roman}, {Filho}, \&
  {S{\'a}nchez Almeida}}]{Trujillo+2017}
{Trujillo}, I., {Roman}, J., {Filho}, M., \& {S{\'a}nchez Almeida}, J. 2017,
  \apj, 836, 191

\bibitem[{{van der Burg} {et~al.}(2016){van der Burg}, {Muzzin}, \&
  {Hoekstra}}]{vanderburg+2016}
{van der Burg}, R. F.~J., {Muzzin}, A., \& {Hoekstra}, H. 2016, \aap, 590, A20

\bibitem[{{van Dokkum} {et~al.}(2016){van Dokkum}, {Abraham}, {Brodie},
  {Conroy}, {Danieli}, {Merritt}, {Mowla}, {Romanowsky}, \& {Zhang}}]{van2016}
{van Dokkum}, P., {Abraham}, R., {Brodie}, J., {et~al.} 2016, \apjl, 828, L6

\bibitem[{{van Dokkum} {et~al.}(2015{\natexlab{a}}){van Dokkum}, {Abraham},
  {Merritt}, {Zhang}, {Geha}, \& {Conroy}}]{vanDokkum+15a}
{van Dokkum}, P.~G., {Abraham}, R., {Merritt}, A., {et~al.} 2015{\natexlab{a}},
  \apjl, 798, L45

\bibitem[{{van Dokkum} {et~al.}(2015{\natexlab{b}}){van Dokkum}, {Romanowsky},
  {Abraham}, {Brodie}, {Conroy}, {Geha}, {Merritt}, {Villaume}, \&
  {Zhang}}]{vanDokkum+15b}
{van Dokkum}, P.~G., {Romanowsky}, A.~J., {Abraham}, R., {et~al.}
  2015{\natexlab{b}}, \apjl, 804, L26

\bibitem[{{Venhola} {et~al.}(2017){Venhola}, {Peletier}, {Laurikainen}, {Salo},
  {Lisker}, {Iodice}, {Capaccioli}, {Verdois Kleijn}, {Valentijn}, {Mieske},
  {Hilker}, {Wittmann}, {van de Ven}, {Grado}, {Spavone}, {Cantiello},
  {Napolitano}, {Paolillo}, \& {Falc{\'o}n-Barroso}}]{Venhola+2017}
{Venhola}, A., {Peletier}, R., {Laurikainen}, E., {et~al.} 2017, \aap, 608,
  A142

\bibitem[{{Villaume} {et~al.}(2022){Villaume}, {Romanowsky}, {Brodie}, {van
  Dokkum}, {Conroy}, {Forbes}, {Danieli}, {Martin}, \&
  {Matuszewski}}]{Villaume+2022}
{Villaume}, A., {Romanowsky}, A.~J., {Brodie}, J., {et~al.} 2022, \apj, 924, 32

\bibitem[{{Wang} {et~al.}(2017){Wang}, {Faber}, {Liu}, {Guo}, {Pacifici},
  {Koo}, {Kassin}, {Mao}, {Fang}, {Chen}, {Koekemoer}, {Kocevski}, \&
  {Ashby}}]{Wang+2017}
{Wang}, W., {Faber}, S.~M., {Liu}, F.~S., {et~al.} 2017, \mnras, 469, 4063

\bibitem[{{Wu} {et~al.}(2005){Wu}, {Shao}, {Mo}, {Xia}, \& {Deng}}]{Wu+2005}
{Wu}, H., {Shao}, Z., {Mo}, H.~J., {Xia}, X., \& {Deng}, Z. 2005, \apj, 622,
  244

\bibitem[{{Yagi} {et~al.}(2016){Yagi}, {Koda}, {Komiyama}, \&
  {Yamanoi}}]{Yagi+2016}
{Yagi}, M., {Koda}, J., {Komiyama}, Y., \& {Yamanoi}, H. 2016, \apjs, 225, 11

\bibitem[{{Zaritsky} {et~al.}(2021){Zaritsky}, {Donnerstein}, {Karunakaran},
  {Barbosa}, {Dey}, {Kadowaki}, {Spekkens}, \& {Zhang}}]{Zaritsky+2021}
{Zaritsky}, D., {Donnerstein}, R., {Karunakaran}, A., {et~al.} 2021, \apjs,
  257, 60

\end{thebibliography}

\end{CJK*}
\end{document}